\documentclass[fleqn,usenatbib]{mnras}

\usepackage{newtxtext,newtxmath}

\usepackage[T1]{fontenc}

\DeclareRobustCommand{\VAN}[3]{#2}
\let\VANthebibliography\thebibliography
\def\thebibliography{\DeclareRobustCommand{\VAN}[3]{##3}\VANthebibliography}

\usepackage{xcolor}
\usepackage{multirow}
\usepackage{soul}

\usepackage{graphicx}	
\usepackage{amsmath}	

\title[Chemical evolution of PSBs]{Chemical evolution of local post-starburst galaxies: Implications for the mass-metallicity relation}

\author[H-H. Leung et al.]{
Ho-Hin Leung,$^{1}$\thanks{E-mail: hhl1@st-andrews.ac.uk}
Vivienne Wild,$^{1}$
Michail Papathomas,$^{2}$
Adam Carnall,$^{3}$
Yirui Zheng,$^{1,4}$\newauthor
Nicholas Boardman,$^{1}$
Cara Wang,$^{1}$
and Peter H. Johansson$^{5}$ 
\\
$^{1}$SUPA\footnote{Scottish Universities Physics Alliance}, School of Physics \& Astronomy, 
University of St Andrews, 
North Haugh, St Andrews, Fife KY16 9SS, UK\\
$^{2}$School of Mathematics and Statistics, University of St Andrews, 
North Haugh, St Andrews, 
Fife, KY16 9SS, UK\\
$^{3}$SUPA Institute for Astronomy, University of Edinburgh, Royal Observatory, Edinburgh EH9 3HJ, UK\\
$^{4}$Department of Astronomy, School of Physics and Astronomy, Shanghai Jiaotong University, Shanghai, 200240, P.R. China \\
$^{5}$Department of Physics, University of Helsinki, Gustaf H\"allstr\"omin katu 2, FI-00014 Helsinki, Finland \\
}

\date{Accepted XXX. Received YYY; in original form ZZZ}

\pubyear{2023}

\begin{document}
\label{firstpage}
\pagerange{\pageref{firstpage}--\pageref{lastpage}}
\maketitle

\begin{abstract}
We use the stellar fossil record to constrain the stellar metallicity evolution and star-formation histories of the post-starburst (PSB) regions within 45 local PSB galaxies from the MaNGA survey. The direct measurement of the regions' stellar metallicity evolution is achieved by a new two-step metallicity model that allows for stellar metallicity to change at the peak of the starburst. We also employ a Gaussian process noise model that accounts for correlated errors introduced by the observational data reduction or inaccuracies in the models. We find that a majority of PSB regions (69\% at $>1\sigma$ significance) increased in stellar metallicity during the recent starburst, with an average increase of $0.8\;$dex and a standard deviation of $0.4\;$dex. A much smaller fraction of PSBs are found to have remained constant (22\%) or declined in metallicity (9\%, average decrease $0.4\;$dex, standard deviation $0.3\;$dex). The pre-burst metallicities of the PSB galaxies are in good agreement with the mass-metallicity (MZ) relation of local star-forming galaxies. These results are consistent with hydrodynamic simulations, which suggest that mergers between gas-rich galaxies are the primary formation mechanism of local PSBs, and rapid metal recycling during the starburst outweighs the impact of dilution by any gas inflows. The final mass-weighted metallicities of the PSB galaxies are consistent with the MZ relation of local passive galaxies. Our results suggest that rapid quenching following a merger-driven starburst is entirely consistent with the observed gap between the stellar mass-metallicity relations of local star-forming and passive galaxies.

\end{abstract}




\section{Introduction} \label{sec:intro}
Since the advent of the first large-scale galaxy surveys such as the 2dF Galaxy Redshift Survey \citep{2df_survey} and the Sloan Digital Sky Survey \citep{SDSS_all}, galaxies have been observed to fall into a bimodal distribution in photometric colours in the local Universe \citep{strateva2001, baldry2004, bell2004, gavazzi2010}. 
The two sub-populations are found to exhibit different distributions across many other properties, including total stellar mass \citep{vulcani2013}, star-formation history (SFH) \citep{kauffmann2003}, kinematics \citep{graham2018}, stellar metallicity \citep{gallazzi2005,peng2015}, radial concentration \citep{hogg2002}, and environment \citep{balogh2004, gavazzi2010}. 
The red sequence consists of quenched, mostly dispersion-dominated galaxies, whilst the blue cloud consists of star-forming, mostly rotationally-supported galaxies. 
The former also have higher stellar metallicity at a given stellar mass than the latter, which can be used to understand the origin of galaxy bimodality by probing the mechanisms of galaxy formation and quenching \citep{peng2015,trussler2020}.

Metallicity is the measurement of the mass of all elements heavier than hydrogen and helium, relative to the total mass of baryons.
The vast majority of metals are produced through stellar processes, including a combination of stellar nucleosynthesis, type Ia and core collapse supernovae (for a review, see \citealt{nomoto2013} and more recently \citealt{maiolino2019}).
These metals are then released into a galaxy's inter-stellar medium (ISM) through mass loss during the red giant phase in lower mass stars ($\approx2-8\mathrm{M_\odot}$) and supernovae in higher mass stars ($\gtrapprox8\mathrm{M_\odot}$). In a closed box system \citep[e.g.][]{tinsley1980} the recycling of this gas into new stars leads to the next generation of stars formed having a higher stellar metallicity than the previous.
However, the closed box model is an unrealistic approximation of galaxies, as interactions with the medium outside the galaxy through inflows and outflows are omitted.

Inflows from the galaxy's circum-galactic medium (CGM) bring in metal-poor gas, diluting the gas reservoir and lowering both the gas-phase and subsequently stellar metallicity. 
Outflows remove gas, slowing down star formation to produce fewer metals. Additionally, outflows that originate from stellar feedback might preferentially remove high metallicity ISM gas from systems, further strengthening the role of outflows in lowering metallicity, particularly in lower mass galaxies \citep{chisholm2018}.
Therefore, the stellar metallicity of a galaxy is a result of the net sum of  three processes: enrichment through stellar processes, inflows, and outflows. These processes are key components of the baryonic cycle in galaxies, which is intrinsically linked to mechanisms that cause galaxy properties to vary with time, including the quenching of star formation.

A key piece of the puzzle to understand the baryonic cycle and the evolution of galaxies is provided by higher redshift galaxy surveys such as UltraVISTA \citep{mccracken2012}.
The surveys found that red quiescent galaxies grow in both number and total stellar mass since $z=4$ \citep{ilbert2013,muzzin2013}, implying star-forming blue cloud galaxies must shut down (quench) their star formation to form quiescent red-sequence galaxies. However, the demographics of red and blue galaxy populations alone are unable to inform on the timescales of these quenching events: the steady growth in quenched galaxies could arise from the average over many individual galaxies with a wide range of different quenching timescales.

As stars form in molecular clouds, the quenching of star formation can be achieved in two ways. The first is the complete consumption of gas following the (likely gradual) termination of the supply of cold gas into the regions of star formation. 
The second is the sudden heating and/or disruption of the molecular clouds due to disruptive events originating from either within or outwith the galaxy.
These two processes are expected to act on different timescales \citep[e.g.][]{schawinski2014}, which is consistent with observational findings that quenching of star formation occurs over varying timescales, ranging from $>5\;$Gyr to $<1\;$Gyr \citep{heavens2004, pacifici2016, bagpipes2018, rowlands2018a}. 

Mechanisms proposed for the slow termination of star formation include natural depletion of gas reservoirs over time through the gradual locking up of gas into stars, the ``maintenance'' of hot gas reservoirs by active galactic nucleus (AGN) feedback preventing cooling of the CGM \citep{croton2006}, 
morphological quenching due to the stabilising force of a central spheroid \citep{martig2009,ceverino2010}, shock heating of higher mass halo gas preventing cooling of gas onto galaxies \citep{dekel2006}, the inhibition of radial inflows of cold gas by the increase in angular momentum of accreted gas due to disc growth \citep{renzini2018,peng2020} and the restriction and/or stripping of galaxy gaseous envelopes by tidal forces in clusters \citep{balogh2000,boselli2006}.

\citet{peng2015} and \citet{trussler2020} have argued that slow quenching mechanisms are the main driver of intermediate and low stellar mass ($M_*<10^{11}M_\odot$) galaxy quenching at $z<1$ due to the higher metallicity of quenched galaxies compared to star-forming galaxies in the local Universe. In this model, the slow decrease in cold gas supply leads to gradual quenching, which allows for star formation to continue with the remaining gas in the galaxy while a lack of continued inflow of low metallicity CGM gas brings reduced dilution effects. The combined effect enhances the metallicity of quenched galaxies with respect to star-forming galaxies. 
\citet{trussler2020} further concluded that, although the decrease in gas supply is the main driver for quenching, a continuous secondary contribution from gas ejection through outflows is required to match the star-formation rates (SFRs) of local passive galaxies particularly at lower stellar masses.

On the other hand, studies that analysed large scale cosmological hydrodynamical simulations have found an important contribution to the build up of the red sequence from rapidly-quenched galaxies (SIMBA, $\approx50$\% contribution of total stellar mass at $z\sim1$: \citealt{montero2019,zheng2022}; IllustrisTNG, $\approx40$\% of galaxies over all redshifts: \citealt{walters2022}). 
Suggested mechanisms that could lead to this rapid quenching of star formation include feedback in the form of violent ejection of gas from the central regions of a galaxy powered by AGN outflows \citep{feruglio2010,cicone2014}. 
Stellar sources such as supernovae and stellar winds could similarly provide substantial feedback, particularly in dense starburst regions \citep{martin1998,martin2005,bolatto2013,molero2023}. 
In clusters, infalling star-forming satellites can experience processes such as ram pressure stripping, thermal evaporation and viscous stripping, which may be powerful enough to remove cold gas directly from star-forming regions \citep{boselli2006}.

Several approaches have been used to measure the relative importance of various quenching mechanisms observationally. 
This includes, but is not limited to, fitting for the SFHs of quiescent galaxies to obtain their quenching timescales \citep[e.g.][]{pacifici2016}, identifying star-forming galaxies with unusually low molecular gas fractions and short depletion times \citep[e.g.][]{gomez-guijarro2022}, and the aforementioned difference in mass-metallicity (MZ) relations between star-forming and quenched galaxies \citep{peng2015,trussler2020}. 
Despite the substantial work in recent years, the various approaches lead to conflicting results in the relative importance of fast and slow quenching mechanisms.

One promising avenue towards resolving this confusion in the literature is the study of post-starburst (PSB) galaxies, which have experienced a recent ($<2\,$Gyr), rapid drop in star formation activity \citep[e.g.][]{wild2020}. 
Studying the prevalence and properties of such objects has the potential to constrain both the contribution of rapid quenching to the growth of the red sequence, as well as the physical mechanisms responsible for such rapid quenching events \citep[e.g.][]{wild2009,rowlands2018b,davis2019,li2019,zheng2020}.
Historically these were first identified as ``E+A'' or ``K+A'' galaxies due to their strong Balmer absorption lines and a lack of nebular emission lines \citep{dressler1983}. 
As a result of their SFH, PSBs exhibit an abundance of A and F type stars, while the shorter-lived O and B stars are largely absent, allowing the pre-burst stellar population to not be heavily outshone \citep[see][for a recent review]{french2021_review}. 
PSBs typically display compact morphologies, in both the local Universe and at higher redshifts \citep[e.g.][]{almaini2017,chen2022}. 
Some studies have suggested that high redshift starburst galaxies such as sub-millimetre galaxies are progenitors of high-redshift PSBs \citep{toft2014,wild2016,wild2020,wilkinson2021} and that low redshift starburst or ultraluminous infrared galaxies (ULIRGs) are progenitors of low-redshift PSBs \citep{hopkins2008,cales2011,french2015,pawlik2018}.

The initial quenching that transitions PSBs away from the starburst phase is expected to be mainly driven by stellar feedback
\citep[see e.g.][]{wild2009}, but current-generation simulations require AGN mechanical feedback (outflows) to completely halt star formation and sustain the reduced SFR after the starburst \citep[e.g.][]{zheng2020}. 
Although PSBs account for only a minor $<1\%$ of the galaxy population at redshift $z\sim0$ \citep{pawlik2016}, the short visibility window of the spectral features means that a considerable fraction of all quenched galaxies could have passed through a PSB phase, particularly at higher redshift \citep{wild2009,wild2016,wild2020,whitaker2012,belli2019,taylor2023}. Therefore, PSBs provide a key testing ground to study the effects of fast quenching mechanisms. 

Measuring the gas-phase metallicity of PSBs is challenging due to the weakness of nebula emission lines and contamination with AGN, shock or diffuse interstellar excitation mechanisms, and can only be achieved in some cases \citep[see][]{rowlands2018b,boardman2024}. However, we might expect substantial chemical evolution to occur during such extreme changes in star formation rate. Given the negative radial metallicity gradients of star forming galaxies \citep[e.g.][]{matteucci1989,zaritsky1994}, the inflow of gas required to drive the centralised starburst common to many PSBs might be expected to pull in lower metallicity gas from the outskirts of the galaxies, reducing metallicity. On the other hand, the very high star formation rates over a short period of time will lead to repeated recycling of gas released from evolved stars and a rapid build up in metals. Given the higher metallicity of quiescent galaxies than star-forming galaxies at given stellar mass \citep{gallazzi2005,peng2015}, which of these processes dominate in reality has important implications for how significantly PSB galaxies, and rapid quenching more generally, can contribute to the build-up of the quiescent population. 

A systematic characterisation of the stellar metallicity evolution of PSBs has not been attempted previously to our knowledge.
In this study, we aim to measure this by taking advantage of the fact that both the pre-burst and starburst stellar population are visible in PSBs' integrated light spectrum. 
To draw a more direct comparison with simulations that focus on the chemical evolution in the cores of starburst galaxies, we focus this study on analysing galaxies with PSB-like centres.
In Section \ref{sec:data}, we describe our data and sample selection. 
In Section \ref{sec:fitting}, we present our method of spectral fitting of the optical continuum through stellar population synthesis models. We test the method with both ``self-consistent'' and simulation-based parameter recovery in Section \ref{sec:tests}, to verify we can recover the SFH and chemical history of PSBs. 
We then apply the method to MaNGA galaxies, present the results in Section \ref{sec:results}, and discuss them in Section \ref{sec:discussion}. 
Where necessary, we assume a cosmology with $\Omega_M=0.3$, $\Omega_\Lambda=0.7$ and $h = 0.7$.
All magnitudes are in the AB system \citep{oke1983}.
We assume a  \citet{kroupa2001} stellar initial mass function (IMF), and take solar metallicity $Z_\odot=0.0142$ \citep{asplund2009}. We re-scale all metallicity measurements quoted from the literature to this solar metallicity for direct comparison. Throughout, we denote lookback time as $t$ and ages of the Universe as $t'$, such that $t'=t_H-t$ where $t_H$ is the age of the Universe.

\section{Data} \label{sec:data}

MaNGA \citep{MANGA} is an integral field spectrograph (IFS) survey of $\approx10000$ $M_* > 10^9 M_\odot$ galaxies (11273 datacubes) in the local $z<0.2$ neighbourhood, a part of the fourth-generation Sloan Digital Sky Survey \citep[SDSS-IV,][]{SDSS_IV} that ran from 2014 to 2020. It used the Sloan Foundation Telescope at Apache Point Observatory \citep{gunn2006} to collect spatially-resolved spectra by using hexagonal bundles of 19 to 127 optical fibres, depending on the apparent sise of the target. The output BOSS spectrographs \citep{smee2013} provide high quality spectra in the wavelength range $3622-10354\text{\normalfont\AA}$ at a spectral resolution of $R\sim2000$\footnote{$R=\lambda/\Delta \lambda_{\mathrm{FWHM}}$}. We access MaNGA data through both the web interface and the python package \texttt{Marvin} \citep{marvin}.

For all MaNGA galaxies in the full data release DR17 \citep{SDSS_DR17}, we obtain redshift from the MaNGA data reduction pipeline \citep{manga_drp,manga_drp_new} and galaxy stellar mass from the NASA-Sloan Atlas \citep[\texttt{NSA\_ELPETRO\_MASS}, a K-correction fit to elliptical Petrosian fluxes, see][]{blanton2011}.
We obtain spectral indices along with other necessary properties from the MaNGA data analysis pipeline \citep{manga_dap,manga_dap2}. 
We adjust the stellar masses from NSA for Hubble constant $h=0.7$. 
Other stellar mass estimates from SDSS-MPA/JHU\footnote{J. Brinchmann: \url{http://www.mpa-garching.mpg.de/SDSS} and \url{http://home.strw.leidenuniv.nl/~jarle/SDSS/}} and the Wisconsin method \citep{chen2012} were also considered, but provided no qualitative changes to the conclusions. 
We limit the sample to $z<0.06$ in favour of local PSBs with good spatial resolution, leaving 7971 galaxies.

Within each MaNGA galaxy's datacube\footnote{The main derived data product of the MaNGA survey; a 3D array with two spatial dimensions and one wavelength dimension. See \cite{manga_drp} for details.}, spaxels\footnote{Spatial pixels, contains the observed spectrum at one particular spatial region on the plane perpendicular to the light of sight.} marked with \texttt{NOCOV}, \texttt{LOWCOV} or \texttt{DONOTUSE} flags are removed. 
To identify PSB spaxels, we broadly follow the methods in \cite{chen2019}, specifically requiring the spaxels' median spectral $\mathrm{SNR} > 8$ per pixel, strength of the $\mathrm{H\delta}$ 
Balmer absorption line after accounting for emission infilling $\mathrm{H\delta_A}>3\text{\normalfont\AA}$ \citep{worthey1997}, equivalent width of the H$\alpha$ nebular emission line after accounting for underlying absorption $\mathrm{W(H\alpha)}<10\text{\normalfont\AA}$\footnote{$\mathrm{W(H\alpha)}$ follows the passband definitions in \cite{manga_dap}, which is a modified version of passbands in \cite{yan2006}.}, and $\log \mathrm{W(H\alpha)}< 0.23\times \mathrm{H\delta_A} - 0.46$. 

To select our galaxy sample, we first selected galaxies with a PSB spaxel fraction $>0.05$ among all classifiable spaxels (spaxels not marked with the previous flags nor the SNR threshold that we impose). Next, we sliced the galaxies into 3 elliptical annuli with $0<R/R_e<0.5$, $0.5<R/R_e<1$ and $1<R/R_e<1.5$, where $R_e$ is the $r$-band elliptical-Petrosian effective radius, using the elliptical polar distance of each spaxel from the galaxy centre. Our galaxy sample is selected to have $>50$\% of the inner annulus spaxels classifiable, and $>50$\% of these spaxels to be classified as a PSB, yielding 54 candidates.
This sample selection is qualitatively similar to the \cite{chen2019} selection of MaNGA galaxies with ``central'' PSB regions.

After the removal of candidates with faulty MaNGA observations (e.g. mismatched redshift and obvious foreground stars, removed 2: 8248-6104, 8601-12703), active galactic nuclei (AGN) broad emission (removed 1: 11004-6104) and datacubes flagged as \texttt{BADFLUX} by the MaNGA DRP (removed 1: 8944-1902, spectrum also appears to be faulty upon visual inspection), the final sample contains 50 PSBs. They span a total stellar mass range of $8.86<\log_{10}M_*/M_\odot<10.94$, as listed in Table \ref{tab:targets} together with other properties.

\begin{table*}
    \centering
    \caption{List of 50 studied PSB galaxies and their properties: 
    (1) MaNGA Plate-IFU identifier;
    (2) MaNGA identifier;
    (3) R.A. (J2000);
    (4) Declination (J2000);
    (5) Redshift;
    (6) $\log_{10}$ total stellar mass fitted from K-corrected elliptical Petrosian photometric fluxes in GALEX/SDSS \textit{FNugriz} bands from the NSA catalogue, adjusted for $h=0.7$;
    (7) Number fraction of classified PSB spaxels among all spaxels not marked with the \texttt{NOCOV} or \texttt{LOWCOV} flags in MaNGA datacubes;
    (8) Final number of stacked spaxels, after excluding spaxels marked with \texttt{DEADFIBER} or \texttt{FORESTAR};
    (9) Mean SNR of the stacked optical spectrum over the full MaNGA wavelength range. 
    The full table is available as supplementary online material.}
    \begin{tabular}{p{1.4cm}p{1.35cm}p{1.5cm}p{1.65cm}p{1.2cm}p{1.1cm}p{1.9cm}p{2.1cm}p{1.6cm}}
    \hline
    Plate-IFU (1) & MaNGA ID (2) & RA (degrees) (3) & Dec. (degrees) (4) & Redshift (5) & $\log_{10}M_*$ ($h=0.7$) (6) & $\frac{\text{PSB spaxels}}{\text{all classifiable spaxels}}$ (7) & Number of stacked spaxels (8) & Stacked mean SNR (9) \\ \hline
    7961-1901   & 1-178035  & 259.53275 & 30.12902  & 0.0296    & 9.99      & 0.45      & 162       & 335.8      \\
    7964-1902   & 1-179682  & 317.42261 & 0.62777   & 0.0242    & 9.73      & 0.07      & 28        & 169.6      \\
    7965-1902   & 1-635485  & 318.50227 & 0.53509   & 0.0269    & 10.41     & 0.91      & 356       & 843.5      \\
    8080-3702   & 1-38062   & 49.22887  & -0.04201  & 0.0231    & 10.19     & 0.39      & 285       & 553.1      \\
    8081-3702   & 1-38166   & 49.94685  & 0.62382   & 0.0247    & 9.45      & 0.12      & 92        & 112.0      \\
    \multicolumn{9}{c}{$\ldots$} \\
    \hline
    \end{tabular}
    \label{tab:targets}
\end{table*}

\begin{figure*}
	\includegraphics[width=\textwidth]{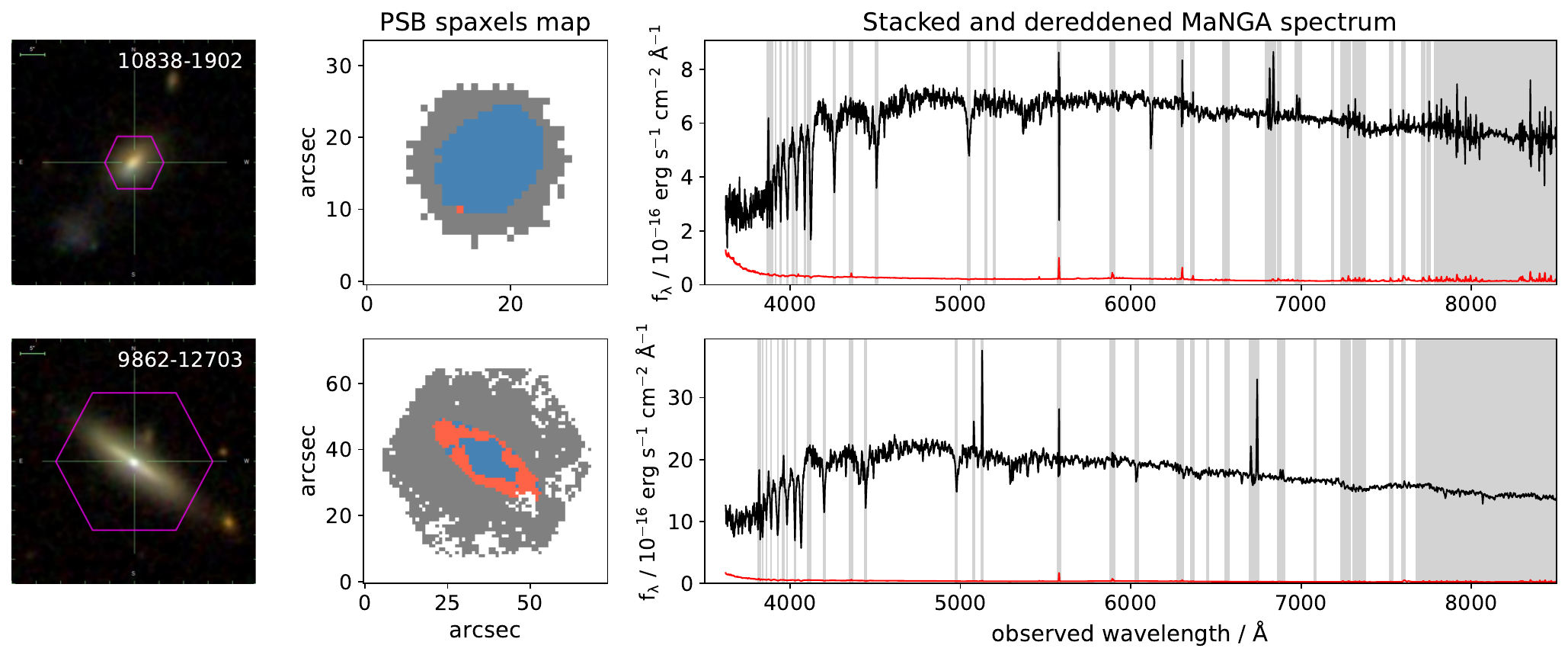}
    \caption{Two typical PSBs from our sample. The top represents PSBs with the vast majority of classifiable spaxels classified as PSB, while the bottom represents PSBs with only a core PSB region. The left panels show the SDSS 3-colour image with the galaxy's Plate-IFU marked on the top right corner. The MaNGA field of view is marked as the pink hexagon. The middle panels show the spaxel selection \citep[broadly following][]{chen2019}, displaying regions with no/faulty observations (transparent), with median spectral $\mathrm{SNR}<8$ too low to be classified (grey), classified as PSB (blue) and classified as non-PSB (red). The right panels show the stacked observed-frame spectrum of the PSB classified spaxels (black), the stacked $1\sigma$ observational uncertainty (red, multiplied by $10\times$ to make visible) and spectral ranges masked during the fitting process (grey bands), including major nebular emission lines, skyline residuals and Balmer infilling. The resulting stacked spectra have a mean SNR of 274 and 482 respectively.}
    \label{fig:masking}
\end{figure*}

We form a stacked PSB-only spectrum for each galaxy, only including spaxel contributions from the PSB-classified spaxels at all radial distances. To ensure spaxel quality, we remove spaxels marked with quality flags \texttt{DEADFIBER} or \texttt{FORESTAR} in MaNGA's H$\alpha$ emission line maps. Spectra are summed unweighted, while uncertainties are summed in quadrature. The stacking of many spaxels for each galaxy allows for a very high SNR to be reached across the full MaNGA wavelength range, with the mean SNR per pixel ranging from $95$ to $>1200$. The SNR of the sample is listed in Table \ref{tab:targets}.
After correcting for Milky Way dust reddening, we further mask major nebular emission lines, residuals of strong skylines (central flux $>5 \times 10^{-16} \ \mathrm{erg\ s^{-1}\ cm^{-2}\ \text{\normalfont\AA}^{-1}}$ in \citealp{skylines}) and Balmer infilling. 
Since \cite{pawlik2018} have previously shown that stellar population synthesis models based on the MILES stellar library \citep{MILES} showed improved recovery of the SFH of PSBs compared to models based on other libraries, we limit the spectra to rest frame $\lambda<7500\text{\normalfont\AA}$ to be fully within the MILES range when fitting. 
Figure \ref{fig:masking} demonstrates the stacking process with two PSBs as examples.

Limiting the spectra to rest frame $\lambda<7500\text{\normalfont\AA}$ potentially loses valuable constraining power on the older stellar population.
Spectral information from longer wavelengths can form a longer wavelength baseline to minimise any age-dust-metallicity degeneracy \citep[see Sections 5.1 and 6.1 of][]{conroy_sed_review}. 
Hence, the flux in the portions with observed frame wavelength $7500<\lambda<9500\text{\normalfont\AA}$ was summed into a single, artificial photometric data point, passed jointly with the trimmed spectra to the fitting framework (Section \ref{sec:fitting}). 
However, no significant differences in the estimated stellar and dust properties were observed with or without the photometric data point, therefore, we limit our analysis to the trimmed spectra for the rest of this study.

\section{Optical continuum spectral fitting}\label{sec:fitting}
To fully utilise the fossil record stored in the high-quality MaNGA spectra, we employ the fully Bayesian spectral energy density (SED) fitting code \textsc{Bagpipes} \citep{bagpipes2018,bagpipes2019}. 
In this section, we describe in detail our spectral fitting procedure of the optical continuum, which includes the assumed parametric SFH model for PSBs from \cite{wild2020} (Section \ref{sec:SFH_model}). 
Motivated by a suite of gas-rich binary major merger simulations that create PSB signatures \citep{zheng2020}, we introduce a novel two-step metallicity model which decouples metallicity before and after the starburst, allowing for any change in stellar metallicity during the starburst to be recovered (Section \ref{sec:two_step}). 
Additionally, we employ a Gaussian process (GP) correlated noise model as an additive term to the physical model's predicted spectrum to account for correlated observational uncertainties and imperfect spectral models (Section \ref{sec:GP_noise}). 
The sampling of the posterior surface is done using the \textsc{MultiNest} nested sampling algorithm \citep{multinest} and its python interface \citep{pymultinest}. As shown in Section \ref{sec:tests} below, our two-step metallicity model also recovers SFH-related parameters more accurately.

\begin{table*}
\centering
	\caption{Model priors used for fitting PSB SEDs. The parameter symbols are described in Sections \ref{sec:fitting} to \ref{sec:GP_noise}, or otherwise have their usual meanings. Some parameters have prior shape $\log_{10}$ uniform, which indicates a flat prior in uniform space $\log(X) \sim U(\log(min), \log(max))$. Redshift is given a uniform prior ranging from 80\% to 120\% of the target's MaNGA redshift ($z$). Note that $\sigma_{\rm disp}$ is not the intrinsic velocity dispersion of the galaxy, as it does not account for the finite resolution of the spectral templates or observational data.}
	\label{tab:priors}
    \begin{tabular}{lllll}
        \hline
        Type & Parameter & Form & Min & Max  \\
        \hline
        SFH & $\log_{10}(M_*/M_\odot)$ & Uniform & 6 & 13 \\
         & $t_{\rm{form}}$ / Gyr & Uniform & 4 & 14 \\
         & $\tau_e$ / Gyr & Uniform & 0.3 & 10 \\
         & $t_{\rm{burst}}$ / Gyr & Uniform & 0 & 4 \\
         & $\alpha$ & $\log_{10}$ Uniform & 0.01 & 1000 \\
         & $\beta$ & Fixed = 250 & - & - \\
         & $f_{\rm{burst}}$ & Uniform & 0 & 1 \\
        Metallicity & $Z_{\rm{old}}/Z_\odot$ & $\log_{10}$ Uniform & 0.014 & 3.52 \\
         & $Z_{\rm{burst}}/Z_\odot$ & $\log_{10}$ Uniform & 0.014 & 3.52 \\
        Dust & $A_V$ / mag & Uniform & 0 & 2 \\
         & birthcloud factor $\eta$ & Uniform & 1 & 5 \\
         & $t_{\rm{birth cloud}}$ / Gyr & Fixed = 0.01 & - & - \\
        GP noise & uncorrelated amplitude $s$ & $\log_{10}$ Uniform & 0.1 & 10 \\
         & correlated amplitude $\sigma$ & $\log_{10}$ Uniform & $10^{-4}$ & 1 \\
         & period/length scale $\rho$ & $\log_{10}$ Uniform & 0.04 & 1.0 \\
         & dampening quality factor $Q$ & Fixed = 0.49 & - & - \\
        Miscellaneous & redshift & Uniform & 0.8 $z$ & 1.2 $z$ \\
         & $\sigma_{\rm{disp}}$ / km/s & $\log_{10}$ Uniform & 40 & 4000 \\ \hline
    \end{tabular}
\end{table*}

Within \textsc{Bagpipes}, we utilise the \cite{bruzual2003} stellar population synthesis models (2016 version), and assume the initial mass function from \cite{kroupa2001}. We apply the two-component dust attenuation law from \cite{wild2007} and \cite{dacunha2008}, with a fixed power-law exponent $n=0.7$ for the interstellar medium (ISM). The dust law asserts that stars younger than $10\,$Myr have a steeper power-law exponent $n=1.3$ and are more attenuated than older stars by a factor $\eta$ \citep[$=1/\mu$ in][]{wild2007,dacunha2008}, as they are assumed to be surrounded by their birth clouds.

Overall, our model has 18 parameters as listed in Table \ref{tab:priors}, 3 fixed and 15 free to be estimated. 
As we follow the Bayesian paradigm, prior distributions are placed on the 15 parameters. 
It is important to also be aware of the imposed prior probability densities on derived physical properties, for example, specific SFR (sSFR) and mass-weighted formation age ($t_\mathrm{M}$), as they can impact the estimated galaxy properties and their uncertainties \citep{carnall2019}. 
These are shown alongside SFH draws from the SFH prior in Figure 3 of \cite{wild2020}.

\subsection{The star-formation history model}\label{sec:SFH_model}
The SFH traces the rate of star formation in a galaxy and all of its progenitors back in time, typically expressed in lookback time. To model both the recent starburst and the underlying older stellar population expected in most local PSBs, we adopt the two-component parametric SFHs of \cite{wild2020}, which provides a good fit to combined spectra and photometry of $z\sim1$ PSBs:
\begin{equation}\label{eq:psb2}
    \mathrm{SFR}(t) \propto \frac{1-f_{\mathrm{burst}}}{\int \psi_e \mathrm{d}t} \times \psi_e(t)\Big|_{t_{\mathrm{form}}>t>t_{\mathrm{burst}}} 
    + \frac{f_{\mathrm{burst}}}{\int \psi_{\mathrm{burst}} \mathrm{d}t} \times \psi_{\mathrm{burst}}(t) \; .
\end{equation}
This is made up of the older, exponential decay component $\psi_e$ and the double power-law starburst component $\psi_{\mathrm{burst}}$, both a function of lookback time $t$. The lookback time when the older population began to form is denoted as $t_{\mathrm{form}}$, while the time since the peak of the starburst is denoted as $t_{\mathrm{burst}}$. The fraction $f_{\mathrm{burst}}$ controls the proportion of mass formed during the starburst. The two components have the forms:
\begin{align}
\label{eq:exp}
    \psi_e(t') &= \exp^{\frac{-t'}{\tau_e}} \\
\label{eq:dpl}
    \psi_{\mathrm{burst}}(t') &= \Big[\big(\frac{t'}{t'_{\mathrm{burst}}}\big)^\alpha 
    + \big(\frac{t'}{t'_{\mathrm{burst}}}\big)^{-\beta}\Big]^{-1}\;.
\end{align}
All times in Equations \ref{eq:exp} and \ref{eq:dpl} are in ages of the Universe, therefore unlike $t_{\mathrm{burst}}$, $t'_{\mathrm{burst}}$ in the starburst component's function represents the age of the Universe at the peak of the starburst. $\tau_e$ is the older population's exponential decay timescale, while $\alpha$ and $\beta$ control the declining and increasing timescales of the burst respectively, with larger values corresponding to steeper slopes. The usage of the fraction $f_{\mathrm{burst}}$ instead of parameterizing the stellar mass formed in the components individually allows for an easier application of a flat prior over $f_{\mathrm{burst}}$. This allows for not only SFH shapes with a strong starburst, but also rapid quenching events of the existing star formation when $f_{\mathrm{burst}}\sim0$.

We investigated allowing the rising starburst slope $\beta$ to vary freely, with a similar prior to $\alpha$. However, parameter recovery tests performed using $\mathrm{SNR}=100$, at the lower end of our observations, showed that $\beta$ is poorly constrained in older starbursts ($t_\mathrm{burst}>1\;$Gyr). Therefore, we fix $\beta=250$, consistent with the typical value found from fits to younger starbursts. 

A common alternative SED fitting method avoids assuming parametric forms for the star-formation history and instead allowing stars to form in fixed or variable bins in time \citep[e.g.][]{starlight2,VESPA,iyer2017,johnson2021}. In general these models do well with smooth SFHs, but are less well suited to galaxies which undergo a rapid change in SFR, due to the need for adaptive variability of the number of time bins. However, both \cite{pawlik2018} and \cite{suess2022} have successfully employed such methods, often referred to as non-parametric, to fit PSBs. \cite{suess2022} increased the number of time bins around the time of the starburst, successfully recovering the rapid rise and fall in SFR of mock PSBs. While this can provide more flexibility in theory, in practice the need to define time bins and in some cases the inclusion of some form of regularisation to smooth between time bins makes the method more model dependent than it first seems. Additionally, no code currently exists which can implement both non parametric SFHs and a GP model to account for correlated noise, which we found crucial for our fitting (see Section \ref{sec:GP_noise}). Therefore, we opt for a parametric SFH approach, noting that the GP noise component is able to account for any slight imperfections in the assumed SFH.

\subsection{Two-step metallicity: insight from PSB merger simulations} \label{sec:two_step}
During integrated light SED fitting, stellar metallicity is often assumed to be constant \citep[e.g.][]{onodera2012,gallazzi2014,bagpipes2018,french2018,wild2020,suess2022}. This is done mainly to limit the dimensionality of the problem, by sacrificing the second-order effects of chemical evolution on observations when compared to that from varying SFH, especially for broad-band photometry. 
This work aims to explore whether this simplification can be removed, and the chemical evolution of PSBs recovered.

To propose a simple yet representative metallicity evolution model for PSBs, we consult the suite of gas-rich binary major merger smoothed-particle hydrodynamics (SPH) simulations that create PSB signatures in \cite{zheng2020}. 
The simulations were performed using the SPH code SPHGal \citep{hu2014,eisenreich2017}, which is an updated version of the Gadget-3 code \citep{springel2005}. 
SPHGal implements sub-resolution astrophysics models from \cite{scannapieco2005,scannapieco2006}, updated by \cite{aumer2013}, and includes gas cooling rates following \cite{wiersma2009}. 
Chemical evolution and stellar feedback from type Ia and type II supernovae, and AGB stars are accounted for \citep[for details, see Section 3.1 of][]{zheng2020}.
The merger progenitor galaxies were set up following \cite{johansson2009} with modifications in the SFR adapted from \cite{lahen2018}, and initial orbital configurations following \cite{naab2003}. The AGN feedback models are from \citet{choi_ena2012} and \citet{choi_ena2014}. The galaxy models have a baryonic particle mass of $1.4\times10^5 \mathrm{M_\odot}$ for both gas and stars, and a gravitational softening length of 28\,pc for all baryonic particles.

\begin{figure*}
	\includegraphics[width=\textwidth]{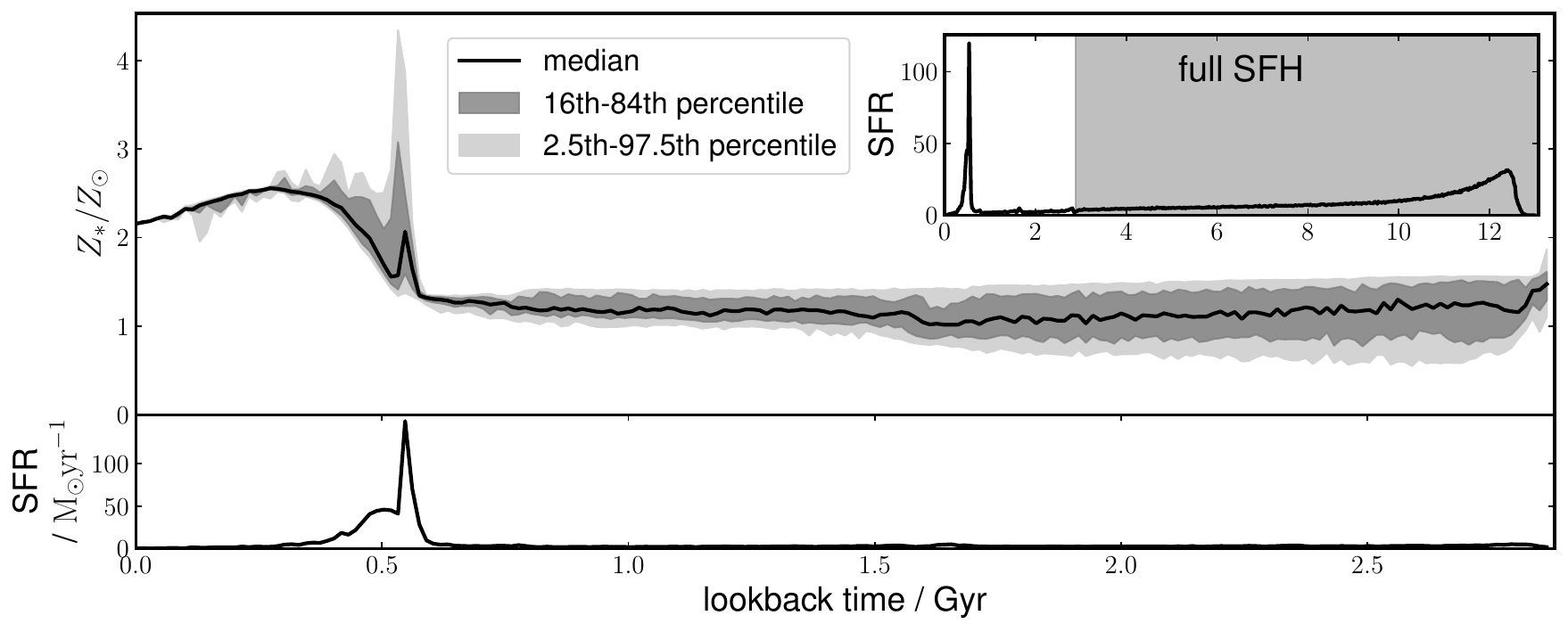}
    \caption{Star-formation history (bottom) and stellar metallicities $(Z_*/Z_\odot)$ of the stars formed (top) in the binary gas-rich major merger simulation 2xSc07 that creates PSB signatures in \protect\cite{zheng2020}. The full SFH is shown in the inset panel, where the shaded region indicates the assumed SFH of the progenitor galaxies. Stellar metallicity increases from $\sim$solar levels to more than twice solar not long after the peak of the starburst at $\sim 550\,$Myr lookback time.}
    \label{fig:yirui_sim}
\end{figure*}

For our fiducial model we use the retrograde-prograde orbit merger simulation of two identical progenitor galaxies with initial gas mass fractions of $f_\mathrm{gas}=0.22$ (2xSc\_07), simulated with mechanical black hole feedback but no radiative feedback, because it results in strong PSB spectral features. 
Figure \ref{fig:yirui_sim} plots the stellar metallicity of simulation particles against their lookback times of formation, together with the simulated SFH. 
When the merger-triggered starburst occurs at $\sim550\,$Myr in lookback time, the newly formed stars have significantly higher stellar metallicity than previous star formation due to rapid recycling of gas to form many generations of stars, and the trend settles on more than twice the pre-burst metallicity after the starburst ends. 
Similar patterns are seen in other gas-rich merger simulations \citep{perez2011,torrey2012}.

We approximate the rapid metallicity increase with a step function and introduce a two-step metallicity model with the time of transition fixed at the peak of the starburst $t_{\mathrm{burst}}$:
\begin{equation}
    Z(t) = \begin{cases}
    Z_{\mathrm{old}} & t>t_{\mathrm{burst}} \\
     Z_{\mathrm{burst}} & t\leq t_{\mathrm{burst}} \; .
    \end{cases}
\end{equation}
Both $t$ and $t_{\mathrm{burst}}$ are in lookback times. The two metallicity levels $Z_{\mathrm{old}}$ and $Z_{\mathrm{burst}}$ are independent and have identical priors, to ensure the model is equally able to fit an increase, decrease or no change in stellar metallicity during the starburst.

We experimented with several more complex metallicity evolution models: a three-step model (pre-burst, during burst, after burst); a gradual increase in metallicity prior to the burst; a two-step metallicity with scatter in the metallicity of coeval stars, following a log-normal or exponential distribution. None provided significantly improved parameter recovery, and given that we do not expect the simulations to be a perfect representation of the real Universe, we felt that any additional model complexity was not justifiable.

\subsection{Treatment of correlated errors} \label{sec:GP_noise}
When fitting photometric data, it is safe to assume the observational uncertainties in the bands are uncorrelated, due to individual photometric bands being observed at different time points, with different instrument set ups. However, when working with spectra consecutive pixels are not independent, due to the many processing steps involved in translating the raw spectroscopic observations into 1D spectral arrays. Following the methods in \citet[][see Section 4 for a detailed discussion regarding the treatment of spectroscopic uncertainties]{bagpipes2019}, we introduce an additive, GP correlated noise component. As well as allowing for correlated uncertainties that stem from the standard data reduction of spectra, this component also serves to account for model-data mismatch that originates from assumptions and approximations involved at all stages of stellar population synthesis: isochrones, stellar spectral templates, SFH, chemical evolution and dust models \citep[see ][for a review]{conroy_sed_review}. 

A GP can be visualised as a series of random variables along one or more continuous axes that represents some physical property. It is a general technique, that has been used to model data in various sub-fields of astronomy, including light curves of X-ray binaries and AGNs \citep{kelly2014}, asteroseismic data of stars \citep{brewer2009,celerite}, exoplanet transits \citep{barclay2015,celerite,chakrabarty2019} and radial velocity measurements \citep{czekala2017}, and the cosmic microwave background \citep{bond1999}. In the case of spectral fitting, the random variables model a series of spectral flux densities along an array of wavelengths, which forms an SED. Each variable is modelled with a Gaussian distribution, such that for a dataset with $N$ values, an N-dimensional Gaussian distribution is constructed. Before the variables are conditioned on the observed data, the prior mean of the Gaussian distributions is typically set as a vector of zeros. This is also adopted in this study. The covariance matrix describes the relationship between each one of the random variables with all other random variables. Each covariance is described by a kernel function that depends on the separation between two observations considering their physical properties. For an in-depth description of GP modelling, see \cite{GPbook}. 

For the fitting of spectra, the GP's covariance matrix allows us to quantify the correlated noise between the measured flux density of any wavelength bin with all other bins. This is useful since it can account for correlated noise on different wavelength scales, where measurements at close-by wavelength bins are expected to correlate more strongly than measurements separated by longer distances. Hence, the close-to-diagonal terms of the covariance matrix will likely have a larger magnitude than off-diagonal terms.

To reduce computational time, we replace the squared exponential kernel used in \cite{bagpipes2019} with a stochastically-driven damped simple harmonic oscillator (SHOTerm), implemented through the \texttt{celerite2} python package \citep{celerite,celerite2}. The GP model of \cite{bagpipes2019} used a covariance matrix describing the covariance between two wavelength bins $j$ and $k$:
\begin{equation}\label{eq:cov_matrix}
    \mathbf{C}_{jk}(\mathbf{\Phi}) = s^2\sigma_j \sigma_k \delta_{jk} + b^2 \exp\left(-\frac{(\lambda_j-\lambda_k)^2}{2l^2}\right) \; ,
\end{equation}
with parameters $\mathbf{\Phi}=(s,b,l)$, where $s$ scales the observational uncertainties $\sigma_{j,k}$ on the SED fluxes, $b$ is the amplitude of the correlated noise and $l$ is the lengthscale of the squared exponential kernel in units of wavelength. $\lambda_j$ and $\lambda_k$ are the wavelengths at indices $j$ and $k$, and $\delta_{jk}$ is the Kronecker delta function. The first term allows for scaling of the uncorrelated input observational noise while the second term is the GP kernel function for correlated noise. In this study, we replace the second term with the \texttt{celerite} SHOTerm kernel function $K$, which is a sum of exponential terms:
\begin{equation} \label{eq:SHOTerm_kernel}
    K_{\mathbf{\alpha}}(|\lambda_j-\lambda_k|) = \sum^{M}_{m=1} a_m \exp\big(-c_m(|\lambda_j-\lambda_k|)\big) \; ,
\end{equation}
where $\mathbf{\alpha}=(\mathbf{a,c})$, with $\mathbf{a}$ and $\mathbf{c}$ vectors with elements $a_m$ and $c_m$ respectively. For a single exponential term of this form, the corresponding inverse covariance matrix is tri-diagonal, which can be computed with a small number of evaluations \citep{rybicki1992,kelly2011}, facilitating a reduction in computation time.

To allow for easier usage of the kernel function, we follow \cite{celerite} to take the Fourier transform of equation (\ref{eq:SHOTerm_kernel}), with the power spectral density
\begin{equation}
    S(\omega) = \sqrt{\frac{2}{\pi}} \frac{S_0 \omega_0^4}{(\omega^2-\omega_0^2)^2 + \omega_0^2\omega^2/Q^2}
\end{equation}
where $\omega_0$ is the frequency of the undamped oscillator, $Q$ is the quality factor of the oscillator, and $S_0$ is proportional to the power of the oscillator at $\omega=\omega_0$: $S(\omega_0)=\sqrt{2/\pi} S_0 Q^2$. To make the function more intuitive, \texttt{celerite2} allows for the option to swap frequency $\omega_0$ with period $\rho$ via the relationship $\rho = 2\pi/\omega_0$, such that period $\rho$ is proportional to a typical lengthscale at which fluxes at $\lambda_j$ and $\lambda_k$ correlate by a standard degree. \texttt{celerite2} also allows for swapping $S_0$ with the standard deviation of the GP realisations $\sigma$ via the relationship $\sigma = \sqrt{S_0\omega_0 Q}$, such that this amplitude parameter is independent of the other parameters $\omega_0$ and $Q$. 

Aiming to emulate the behaviour of the squared exponential kernel, which works well on spectral fitting problems, we match its auto-correlation curves with those from the SHOTerm kernel. This process is described in Appendix \ref{apx:GP}. This replacement of kernels allowed for a $\sim100$ fold reduction in computational time.

\section{Testing of fitting methods}\label{sec:tests}
In this section we demonstrate that the combination of the 2-component SFH model with the two-step metallicity model can recover relevant galaxy parameters when presented with spectra of PSBs, with low systematic biases. We perform two types of parameter recovery tests: a ``self-consistent” test is described in Section \ref{sec:self-consistant}, and a smooth particle hydrodynamic (SPH) simulation test is described in Section \ref{sec:particle_test}.

\subsection{Self-consistent parameter recovery}\label{sec:self-consistant}
\begin{figure*}
    \centering
    \includegraphics[width=0.93\textwidth]{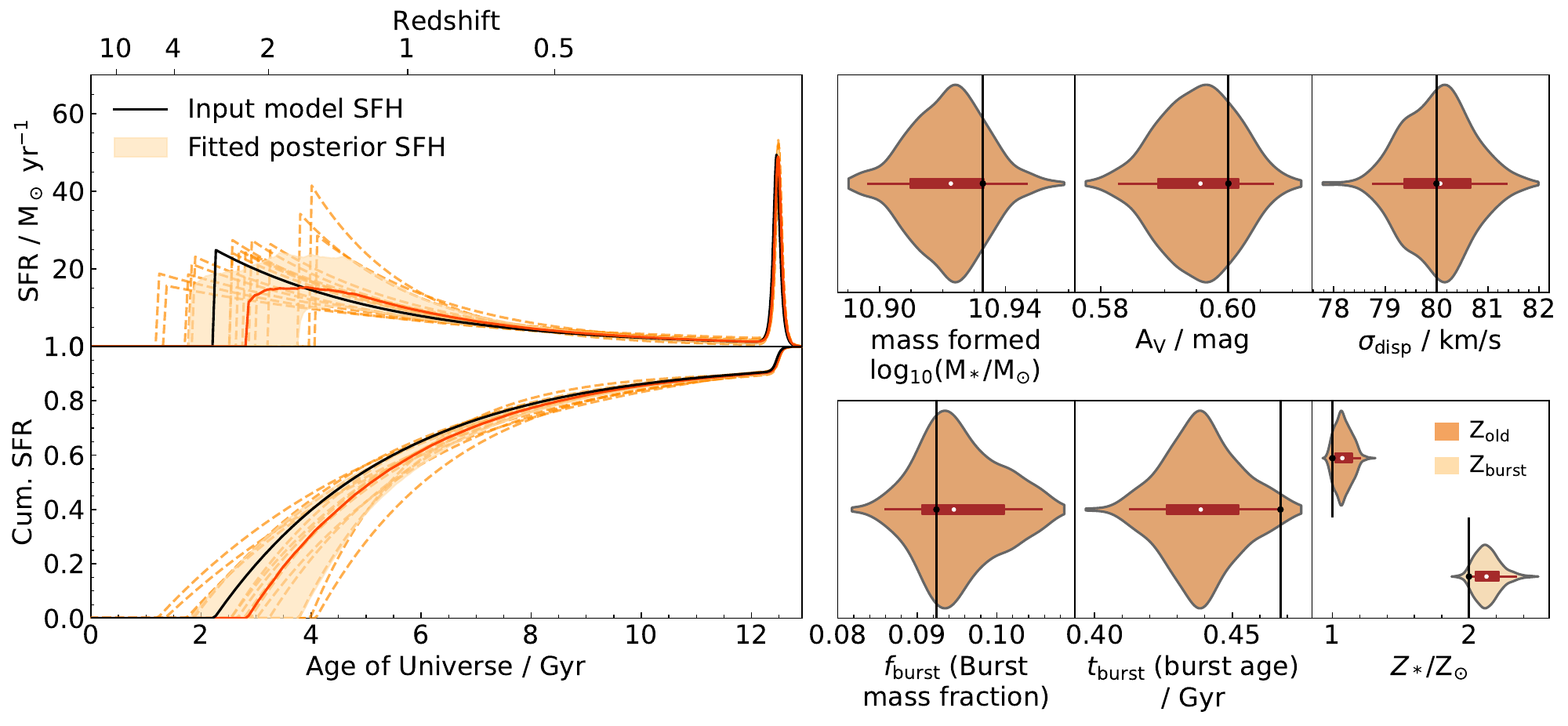}
    \caption{Self-consistent parameter recovery test with the two-step metallicity model. \textbf{Left}: SFH (top) and fractional cumulative SFH (bottom), showing the input truth (black line), posterior median (solid orange line) and its $1\sigma$ region (shaded), and 15 random draws from the posterior fit (dashed lines). \textbf{Right}: Violin plots showing posterior distributions of total stellar mass formed ($\log_{10}M_*/M_\odot$), extinction in V band ($A_V$), velocity dispersion ($\sigma_\mathrm{disp}$), burst mass fraction ($f_\mathrm{burst}$), age of the burst ($t_\mathrm{burst}$) and metallicity levels. The height corresponds to the distribution's density, while the central box plot marks its median (white dot), $1\sigma$ region (thick brown bar) and $2\sigma$ region (thin brown line). The vertical black lines indicate the input truths. In the lower right panel, the lighter and darker shaded violins correspond to the posterior of the burst and older metallicities, respectively. All parameters are estimated with accuracy within $2\sigma$, and the metallicity change is recovered.}
    \label{fig:self-consistant_test}
\end{figure*}

The ``self-consistent” test involves generating mock PSB spectra using the functional forms for all parameters, including SFH, metallicity evolution and dust, then fitting the mock spectra with the same functional forms and spectral templates used for mock spectra generation. This setup ensures there is no model-data mismatch nor correlated errors. If the parameter recovery is successful across a large range of input values, it indicates the fitting process can recover the required parameters with minimal degeneracies when the model can perfectly describe the data.

We generate spectra using \textsc{Bagpipes}, with identical wavelength range and spectral resolution as our real MaNGA spectra, perturbing the generated spectra with Gaussian uncorrelated errors assuming $\mathrm{SNR}=100$ similar to the minimum SNR of our observed spectra. Typical dust and dispersion values were assumed, based on the results from our observed sample ($A_V=0.6$, $\eta=3$, $\sigma_\mathrm{disp}=80\;$km/s). Since we do not inject correlated errors, there is no need to include the GP noise component during fitting.

Figure \ref{fig:self-consistant_test} shows the recovery performance of a self-consistent test with mock input parameter values similar to the SPH-simulated PSB in Figure \ref{fig:yirui_sim}, using the two-step metallicity model. The left panels demonstrate that we are able to recover the input SFH to within $1\sigma$ for nearly all lookback times. In the top left panel, the apparent mismatch between the posterior median SFH (solid orange) and input SFH (solid black) before $z=1$ is partly a plotting artefact. Since each posterior sample is an exponential decay function with an abrupt increase in SFR at $t_\mathrm{form}$, the median SFR includes a steadily decreasing fraction of models with no star formation. Hence we also plot the SFHs of 15 posterior samples in the same panel, and show the cumulative fraction of the total stellar mass formed against the age of the Universe in the bottom left panel as an alternative visualisation. 
In the cumulative SFH, it is easier to see that the discrepancy between the fitted median and input curves is $<1\sigma$.

The right panels of Figure \ref{fig:self-consistant_test} show violin representations of the posterior distributions for seven key parameters, demonstrating they are recovered to within $2\sigma$ of the input truths (solid bars). Particularly, we are able to recover the difference between the pre-burst and starburst stellar metallicities, with a high degree of confidence.

\subsubsection{Sample of self-consistent model tests}
To understand whether the offsets observed between true values posterior median estimates in Figure \ref{fig:self-consistant_test} are systematic, we repeated the recovery test 100 times with randomly drawn input values\footnote{The total stellar mass formed and redshift are fixed, as these do not alter the shape of the spectrum, and varying them will not provide additional insight.} from the priors in Table \ref{tab:priors}. Variations in dust and velocity dispersion are omitted due to computational time limitations (although see below for a comparison when these are included). We only fit mock spectra that are classified as PSBs under our selection criteria using $\mathrm{H\delta_A}$ and $\mathrm{W(H\alpha)}$ (Section \ref{sec:data}). The mean and standard deviation of the offset (median of posterior $-$ input truth), and fitting uncertainty for all tests are listed in Table \ref{tab:self-consistent_bias}. Identifying parameters where the mean offset is greater than the mean uncertainty, we find a very slight average overestimation in burst age. However, this is two orders of magnitude smaller than the range of our sample's estimated burst ages (Section \ref{sec:results}), thus this does not impact our main results. We verify that there is no bias for finding a change in metallicity when the true metallicity remains constant, and incorrectly assigning an increase in metallicity when the true metallicity falls, and vice versa.

\begin{table}
    \centering
    \caption{The mean and standard deviation of offsets ($\Delta$, median of posterior $-$ input truth), and mean uncertainties from 100 self-consistent parameter recovery tests using the two-step metallicity model. The input values are randomly drawn from the priors given in Table \ref{tab:priors}, but were then checked to ensure the resulting spectrum satisfied our PSB selection criteria. We list here the mean offset and fitting uncertainty averaged across all 100 tests. All symbols follow the definitions in Section \ref{sec:fitting}.}
    \begin{tabular}{llll}
        \hline
        \multirow{2}{*}{Parameter} & \multirow{2}{*}{Mean offset ($\overline{\Delta}$)} & \multirow{2}{*}{SD of offset} & Mean $1\sigma$ \\
        & & & uncertainty \\
        \hline
        $\mathrm{log_{10}(M_*/M_{\odot})}$ & -0.002 & 0.026 & 0.011  \\
        $t_\mathrm{form}$ / Gyr & 0.19 & 1.95 & 1.33  \\
        $t_\mathrm{burst}$ / Gyr & 0.049 & 0.165 & 0.038  \\
        $f_\mathrm{burst}$ & 0.016 & 0.076 & 0.022  \\
        $Z_{\mathrm{old}}\ /\ \mathrm{Z_{\odot}}$ & -0.001 & 0.084 & 0.019  \\
        $Z_{\mathrm{burst}}\ /\ \mathrm{Z_{\odot}}$ & -0.023 & 0.197 & 0.038  \\
        \hline
    \end{tabular}
    \label{tab:self-consistent_bias}
\end{table}

In addition to the test shown in Figure \ref{fig:self-consistant_test}, five self-consistent parameter recovery tests with dust and velocity dispersion are performed based on randomly drawn input values, including $A_V$, $\eta$ and $\sigma_\mathrm{disp}$. 
Comparing the five test results to the 100 above that did not have the dust component and added velocity dispersion, a $\sim40\%$ increase in estimation uncertainty is seen across individual SFH properties and metallicity. 
Despite this, with dust and velocity dispersion, the recovered values remain within $2\sigma$ of the input truths (see Figure \ref{fig:self-consistant_test}).

These tests show that, in the absence of model-data mismatch and correlated noise, we can recover the input parameters of the two-step metallicity model, via integrated light MaNGA-like spectra, for a wide range of PSBs with varying stellar and metallicity histories.

\subsection{SPH Simulation parameter recovery} \label{sec:particle_test}

\begin{figure*}
    \centering
    \includegraphics[width=0.93\textwidth]{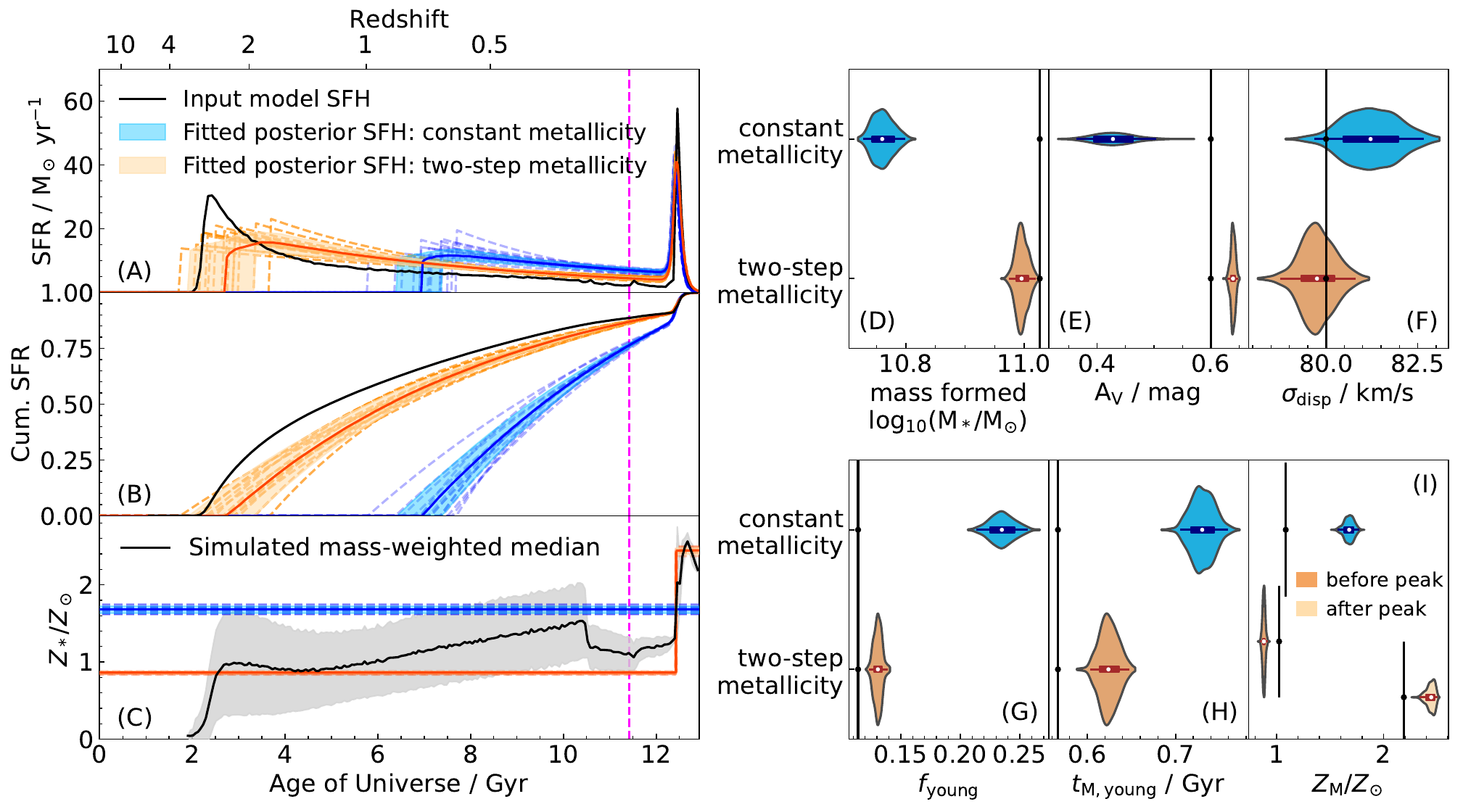}
    \caption{Simulation-based parameter recovery test, comparing the results of fitting the constant (blue) and two-step (orange) metallicity models to star particles generated from an SPH merger simulation \citep[from][see Figure \ref{fig:yirui_sim}]{zheng2020}. Most panels are as per Figure \ref{fig:self-consistant_test}. Panel (C) additionally presents the stellar metallicity evolution; lines and shaded regions have the same meaning as for the SFH panels. The vertical magenta line marks a lookback time of $1.5\;$Gyr, before and after which we calculate the fraction of mass formed within $t<1.5\;$Gyr ($f_\mathrm{young}$) and the mass-weighted mean stellar age within $t<1.5\;$Gyr ($t_\mathrm{M,young}$) shown in panels (G) and (H). In panel (I), the top-most black vertical line indicates the mass-weighted metallicity throughout the simulation, while the following two are the mass-weighted metallicity of stars formed before and after the peak of the starburst. The two-step metallicity model out-performs the constant model in the recovery of all parameters.}
    \label{fig:particle_test}
\end{figure*}

\begin{table*}
    \centering
    \caption{The offsets ($\Delta$, median of posterior - input truth) from 46 parameter recovery tests fitting SPH-derived PSB spectra with the constant and two-step metallicity models. We list here the mean and standard deviation of offsets and fitting uncertainty averaged across all 46 simulated spectra. Parameters (1), (7) and (8) follow the meanings in Section \ref{sec:fitting}, while the other parameters are 
    (2) mass-weighted mean stellar age before lookback time $t=1.5\;$Gyr,
    (3) mass-weighted mean stellar age after lookback time $t=1.5\;$Gyr,
    (4) fraction of mass formed within $t<1.5\;$Gyr,
    (5) mass-weighted mean stellar metallicity before the corresponding simulation's peak SFR of the starburst,
    and (6) mass-weighted mean stellar metallicity after the corresponding simulation's peak SFR of the starburst. For a wide range of SFHs, metallicity evolution and dust properties, the two-step metallicity model shows smaller offsets than the constant model.}
    \begin{tabular}{lllllll}
        \hline
          & \multicolumn{3}{c}{Constant metallicity model} & \multicolumn{3}{c}{Two-step metallicity model} \\ \cline{2-7} 
        Parameter & Mean offset ($\overline{\Delta}$) & SD of offset & Mean $1\sigma$ uncertainty & Mean offset ($\overline{\Delta}$) & SD of offset & Mean $1\sigma$ uncertainty \\
        \hline
        $\mathrm{log_{10}(M_*/M_{\odot})}$ (1) & -0.186 & 0.054 & 0.013 & -0.048 & 0.028 & 0.012  \\
        $t_\mathrm{M,old}\ /\ \mathrm{Gyr}$ (2) & -3.26 & 0.70 & 0.19 & -1.55 & 0.87 & 0.34  \\
        $t_\mathrm{M,young}\ /\ \mathrm{Gyr}$ (3) & 0.153 & 0.052 & 0.009 & 0.027 & 0.073 & 0.015  \\
        $f_{\mathrm{young}}$ (4) & 0.108 & 0.043 & 0.008 & 0.019 & 0.019 & 0.005  \\
        $Z_\mathrm{M,before\ peak}\ /\ \mathrm{Z_{\odot}}$ (5) & 0.464 & 0.280 & 0.041 & 0.007 & 0.263 & 0.020  \\
        $Z_\mathrm{M,after\ peak}\ /\ \mathrm{Z_{\odot}}$ (6) & -0.939 & 0.268 & 0.041 & -0.277 & 0.668 & 0.040  \\
        $\mathrm{A_V}$ / mag (7) & -0.068 & 0.086 & 0.021 & 0.025 & 0.017 & 0.009  \\
        $\sigma_{disp}$ / km/s (8) & 1.47 & 1.05 & 0.65 & 0.19 & 0.94 & 0.59  \\
        \hline
    \end{tabular}
    \label{tab:particle_bias}
\end{table*}

The second parameter recovery test involved generating mock PSB spectra from the stellar particles of the SPH simulations in \cite{zheng2020}, and fitting them with our assumed models to see whether we can recover the underlying galaxy properties. Unlike in the ``self-consistent" tests above, the star formation and chemical evolution history of the SPH simulations is complex, and cannot be perfectly described by the simple functional forms of our model. Additionally, stars formed coevally in the simulation can have a range of metallicities, which is not possible in our model. Thus, the mock spectra are created from galaxy properties that do not exist within the prior model space, so parameters can only be approximately recovered. Any inaccuracies and biases found during this test allows for conclusions to be drawn concerning the models' performances when tasked with real data, which will exhibit similar characteristics. 

While investigating the cause of the small number of self-consistent parameter recovery tests with large discrepancies between estimated and true values, we discovered that many occur as the mock galaxy's $t_\mathrm{form}$ approaches the age of the Universe. The rate of change in the flux of a galaxy spectrum with time decreases with increasing stellar age, hence, errors on $t_\mathrm{form}$ increase for the oldest galaxies. This issue is not due to the changing metallicity, as it is seen in the parameter recovery tests of both constant and two-step metallicity models. Unfortunately, all the SPH simulations in \citet{zheng2020} were initialised with analytic templates that began their star formation at age of the Universe $t=0.5\,$Gyr. Therefore, to enable a better insight into the recovery of star formation and metallicity histories in PSBs, we scale the age of all simulated stellar particles down by 15\%, preserving the total stellar mass formed and the shape and chronology of the SFH. 
The shift away from simulated SFHs that began at very low age of the Universe does not impact our results, as the typical estimated age of the Universe when star formation began for our sample ($t'_\mathrm{form}$) is $>3\;$Gyr. 
Mock spectra are generated exactly as described in Section \ref{sec:self-consistant}, with the same dust properties, velocity dispersion, SNR and uncorrelated Gaussian noise. 
Due to model-data mismatch caused by the simulations' parameters lying outside of the prior space, the GP noise component is included when fitting.

Figure \ref{fig:particle_test} compares the results of fitting spectra constructed from the binary merger simulation shown in Figure \ref{fig:yirui_sim} with the constant and two-step metallicity models. Due to the model no-longer existing within the model prior space, we no longer expect perfect parameter recovery. The top left panels show that the two-step metallicity model outperforms the constant model when recovering the SFH of simulated PSBs that underwent changes in stellar metallicity. The bottom left panel shows the fitted two-step metallicity model closely follows the simulation's metallicity evolution\footnote{The sudden drop in metallicity of the simulation at 10.5 Gyr is a result of the switch from analytic progenitor galaxies to SPH simulation and is not physical.}.

As there is no direct input ``truth” corresponding to many of the fitted parameters, in the right hand violin plots we instead compare the fraction of mass formed within $t<1.5\;$Gyr ($f_\mathrm{young}$), mass-weighted mean stellar age within lookback time $t<1.5\;$Gyr ($t_\mathrm{M,young}$), and the mass-weighted mean stellar metallicity throughout the entire simulation, as well as before and after the peak SFR of the starburst. In all cases, the two-step metallicity model substantially outperforms the constant metallicity model in recovering the underlying galaxy properties.

In the bottom right panel, we see that the fitted metallicity of the constant metallicity model is $>5\sigma$ higher than the true overall mass-weighted metallicity. The over-estimation of the older stellar population's metallicity results in a redder old-star spectrum, leading to a younger fitted $t_\mathrm{form}$. The failure to recover the light from old stars (formed before $6\;$Gyr), leads to an underestimation of total stellar mass formed by $>0.2\;$dex. On the other hand, the under-estimation of the burst population's metallicity results in a bluer young-star spectrum, leading to an overestimation of the burst age to compensate. The flexibility of the two metallicity levels allows for these problems to be mitigated. As a result, the violin plots show a significantly more accurate recovery of all parameters displayed. 

To verify that the two-step metallicity model also enables good recovery of input true values when metallicity declines during the starburst, we artificially flip the metallicity of stellar particles in the simulation, to simulate a decrease in metallicity. We found the two-step model again results in a superior recovery of the SFH compared to the constant model.

\subsubsection{Sample of simulation recovery tests}
To investigate the possible bias on recovered parameters, we expand the simulation parameter recovery test to a suite of 46 tests performed on PSB spectra predicted from the simulations in \cite{zheng2020}. All tests assumed the same dust properties, velocity dispersion, SNR and perturbation of the generated spectrum as previous tests. We only use the simulation runs that resulted in an obvious starburst followed by rapid quenching i.e. prograde-prograde and retrograde-prograde simulations, more gas rich progenitors, and those with mechanical black hole feedback, but without radiative feedback. The inclusion of radiative feedback was found to be too effective at suppressing the increased star formation after the merger, leading to no/very weak PSB signatures in the resulting galaxy \citep[see][]{zheng2020}. 
Specifically, these were simulations Sa\_Sc\_00, Sa\_Sd\_00, 2xSc\_00, Sc\_Sd\_00, 2xSd\_00, Sa\_Sc\_07, Sa\_Sd\_07, 2xSc\_07, Sc\_Sd\_07 and 2xSd\_07. The initial gas mass fractions of the progenitors are 0.17, 0.22 and 0.31 for Sa, Sc and Sd, respectively.

From each of the 10 SPH simulations, we extract 10 post-burst spectra equally spaced in time from the peak of the starburst to the end of the simulation. 
We do this by discarding star particles formed after each time point, input the remaining particles’ stellar metallicity and ages (shifted to the new time of observation), into \textsc{Bagpipes} which then constructs the integrated spectrum from SSPs in the same way as our models. 
We then measure the $\mathrm{H\delta_A}$ and $\mathrm{W(H\alpha)}$ from the integrated spectra, and check whether they pass our selection criteria (Section \ref{sec:data}). This results in 46 simulated PSB spectra at $0.11-0.71\;$Gyr since the peak SFR of the starburst.
Similar to the example shown in Figure \ref{fig:yirui_sim}, all chosen simulations exhibit rapid stellar metallicity increase during the starburst, leading to a much higher recent ($t<1.5\;$Gyr) mass-weighted metallicity than before the starburst ($t>1.5\;$Gyr).

The mean offset and fitted uncertainty for both constant and two-step metallicity models are presented in Table \ref{tab:particle_bias}. The two-step metallicity model achieves less bias in SFH-related parameters (total mass formed, $t_\mathrm{M,young}$ and $f_\mathrm{young}$), $A_V$ and $\sigma_\mathrm{disp}$ than the constant model, for a wide range of PSBs with varying SFHs and chemical evolution, ages and scatter. The two-step metallicity model returns a small mean offset in both metallicity measurements, which indicates the model is able to accurately recover the metallicity change for a broad range of simulated PSBs.

Tables \ref{tab:self-consistent_bias} and \ref{tab:particle_bias} demonstrate that the standard deviation of the offset between true and fitted parameters is greater than their respective mean $1\sigma$ uncertainty. This is expected, due to the model being an imperfect representation of the data, and illustrates why mock recovery tests with semi-realistic data are necessary. We are careful to take account of this in the interpretation of our results below, but also note that the dynamic range of the parameters of interest is significantly larger than even the standard deviation of the offset in the mock recovery tests.

We note that among the suite of simulation recovery tests there are several outliers with larger offsets in metallicity estimated with the two-step metallicity model than with the constant model. These are limited to models with a starburst peaking recently ($0.2-0.4\;$Gyr in lookback time). For these to be selected as PSBs, they have a correspondingly rapid quenching time (e-folding timescale $\tau\sim 50\;$Myr). In this case, the two-step metallicity model can suffer from a degeneracy between the true solution and a slightly older starburst with higher burst mass, longer quenching timescales and a declining stellar metallicity. In most cases where the two-step model fails to recover the correct metallicity evolution, the constant model also suffers from a similar older, more massive starburst degeneracy, albeit to a less severe degree. PSBs with such recent starbursts are not found in our observed sample (Table \ref{tab:results}). Therefore, we do not expect this to be a significant concern for our results.

\begin{figure*}
    \centering
    \includegraphics[width=\textwidth]{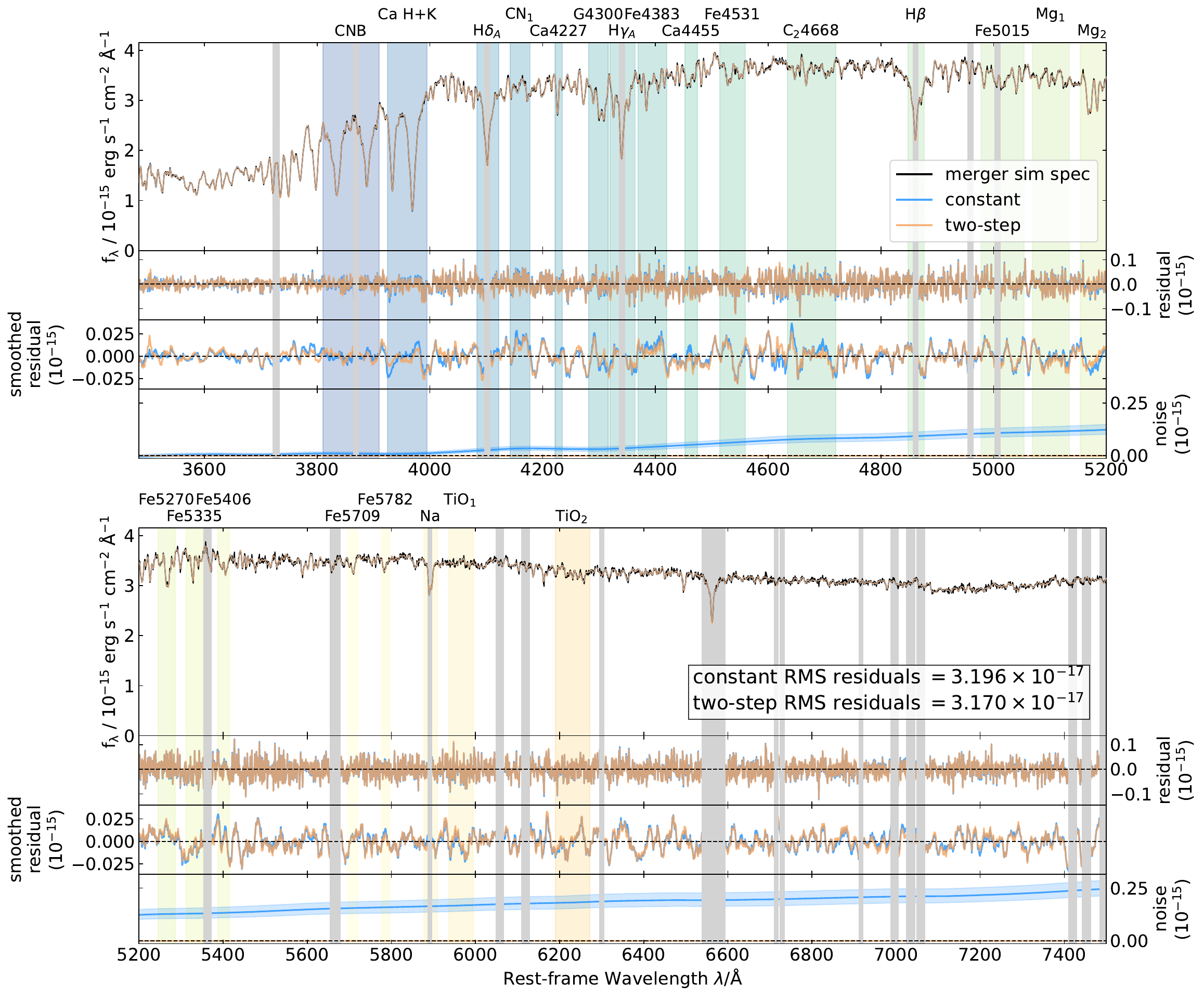}
    \caption{Comparing the spectral fitting performance of the constant (blue) and the two-step (orange) metallicity models applied to a mock PSB spectrum generated from an SPH merger simulation (Figure \ref{fig:particle_test}). The spectrum is split into two rows for clarity. In each row, the main panel shows the input mock spectrum (black) and the best-fit spectra from the two metallicity models (blue and orange). The lower panels show the fitting residuals, residuals smoothed by a mean boxcar of width 11 pixels, the GP noise contribution to each metallicity models' best-fit spectral model, respectively. The orange GP noise curve has a much smaller amplitude compared to the blue curve and largely lies along the black dashed line. All y-axes have the same units. The vertical grey bars indicate masked regions where emission lines would appear in the MaNGA data. Coloured bars mark the wavelengths of common spectral indices. Root mean square residuals of the maximum likelihood posterior sample spectrum (including GP noise) for both fits are also shown. The two-step model achieves a better fit of the SPH PSB mock, particularly in the spectral regions of the calcium H+K lines and iron and magnesium metallicity indices.}
    \label{fig:particle_test_spec}
\end{figure*}

To understand the spectral features that drive the better parameter recovery by the two-step metallicity model, Figure \ref{fig:particle_test_spec} compares the fitted spectra from the two metallicity models. The lower three plots in each panel of Figure \ref{fig:particle_test_spec} show the fitting residuals, the residuals smoothed by a mean boxcar of width 11 pixels and the fitted spectra's GP noise contributions. Vertical coloured regions mark the wavelengths of the Lick indices \citep{worthey1994,worthey1997} and indices CNB and H+K from \cite{brodie1986}. 
The root mean square residual for both models is noted in the lower panels.

At first sight the fitted spectra from both metallicity models appear to be very well matched with the mock spectrum.
However, the smoothed residual reveals differences at all wavelengths. 
The two-step model's smoothed residual is smaller particularly within the calcium H+K lines and the iron and magnesium metallicity indices, which contributes to the two-step model's slightly lower root mean square residual. 
This is consistent with the better fit to stellar metallicities obtained when using the two-step model. 

The constant metallicity model's GP noise component led to a best fit model with a prominent slope (the red-end is $2.51^{+0.38}_{-0.36} \times 10^{-16}\; \mathrm{erg\; s^{-1}\; cm^{-2}\; \text{\normalfont\AA}^{-1}}$ higher than the blue-end), while the physical spectrum is significantly bluer than the input mock spectrum.
This could arise from a combination of incorrectly estimated dust attenuation curve, incorrect metallicity or SFH properties, leading to the larger offsets for all properties in Table \ref{tab:particle_bias}.
The two-step metallicity model's GP noise component has a much smaller amplitude at all wavelengths (amplitude $RMS=4.2^{+8.4}_{-2.8} \times 10^{-19}\; \mathrm{erg\; s^{-1}\; cm^{-2}\; \text{\normalfont\AA}^{-1}}$, small enough to be hidden behind the dashed line), and no overall slope. 
This indicates that the two-step model's higher degree of flexibility allows for a better approximation of the mock spectrum generated from simulation, and only minor corrections from the GP noise are required.
For completeness, in Appendix \ref{apx:noGP_test} we include a SPH simulation recovery test without the use of the GP noise component to investigate the impact of the component's correction particularly for the constant metallicity model's fit. Qualitatively similar conclusions were made.

To summarise, the two-step metallicity model allows for metallicity evolution during the starburst to be traced without significant estimation biases or reduction in fitting precision due to the increased complexity, while allowing for better recovery of the SFH and its parameters of the galaxy.

\section{Results}\label{sec:results}
We fit all 50 PSBs with the two-step metallicity model described in Section \ref{sec:fitting}, including the new GP correlated noise model. 
5/50 ($10\%$) resulted in fitted GP noise components showing obvious trends across the fitted spectral range, with amplitudes much larger than the observational uncertainty scaled by the posterior median uncorrelated noise amplitude ($s$ in Equation \ref{eq:cov_matrix}). 
This can potentially indicate additional sources, such as AGN or foreground/background sources in their fitted spectra, or complex stellar/dust properties that cannot be adequately fit with the model. 
Due to these considerations, their results are excluded from further analysis (Plate-IFU: 7965-1902, 8080-3702, 9494-3701, 9495-3702, 9507-12704). 
All remaining 45 were found to have clear PSB SFHs, where rapid quenching follows an episode of strong starburst. 
In general, the PSB regions underwent a starburst $\sim 1\;$Gyr before observation, with the youngest starburst occurring $\approx 0.45\;$Gyr ago. The starbursts have a wide range in mass fraction ($\approx0.06-0.94$).
The fitted SFH properties, metallicity levels, $A_V$ and reduced chi-squared\footnote{Reduced chi-squared $\chi^2_\nu = \big (\sum_i\frac{{(O_i-C_i)^2}}{s^2\sigma^2_i}\big)/\nu$ where $O_i$ is the observed spectrum, $C_i$ is the maximum likelihood posterior sample spectrum including the GP noise, and $\nu$ is the degree of freedom. Observational uncertainty $\sigma_i$ is multiplied by the median posterior uncorrelated scaling factor $s$.} of the maximum likelihood posterior sample spectrum are reported in Table \ref{tab:results} and discussed in the following sections. All fitted SFHs and metallicity evolution are plotted in Appendix \ref{apx:SFH}. 

\begin{figure*}
    \centering
    \includegraphics[width=\textwidth]{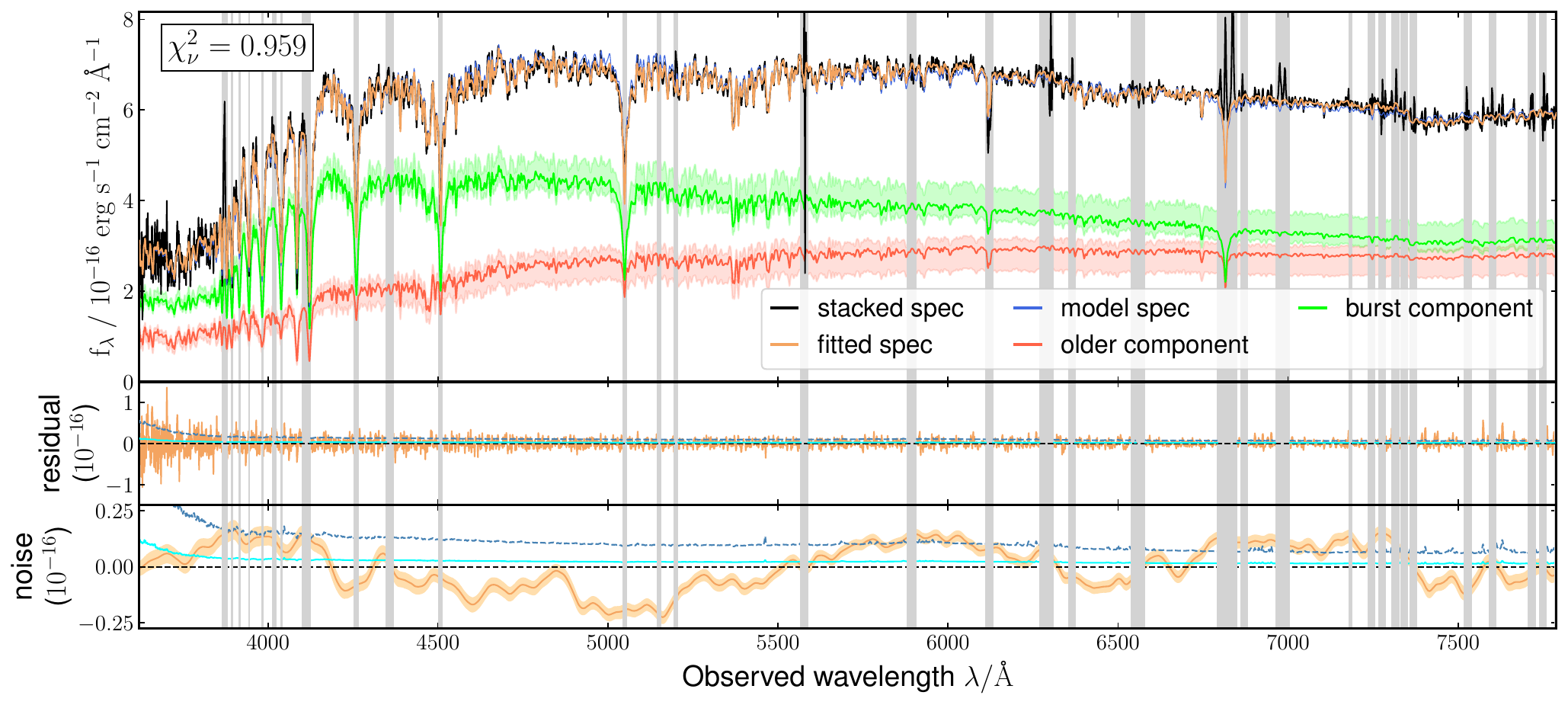}
    \caption{Example of a fitted stacked MaNGA spectrum and the contributing components (10838-1902, the top galaxy in Figure \ref{fig:masking}). \textbf{Top}: The stacked observed spectrum created by combining only spaxels classified as PSB (black), the posterior best fit spectrum and its distribution (orange line and orange shaded region), which includes contribution both from the physical model and GP noise. The physical model (blue) can be separated to show light contributions from the starburst (lime and lime shaded region) and the older components (red and red shaded region). The reduced chi-squared value of the maximum likelihood posterior sample spectrum (including GP noise), $\chi^2_\nu$, is shown. \textbf{Middle}: The fitting residual (orange), defined as the observed stacked spectrum (black curve) - posterior best-fit spectrum (orange curve). The light blue line and the blue dashed line show the input observational uncertainty before and after scaling by the fitted noise scaling factor $s$, respectively. An increase of around $\times3-5$ is typically required. \textbf{Bottom}: The fitted GP noise component and its distribution in orange, with blue curves as above. The majority of the fitted GP noise flux lies below the scaled observational uncertainty (blue dashed) and there is no obvious global trend, thus, this galaxy is recognised as a good fit. Note that y-axes have the same units, but the three panels vary in scaling. In all panels, vertical grey bands indicate regions masked due to skyline residuals, strong nebular emission lines or Balmer infilling.}
    \label{fig:fit_example}
\end{figure*}

Figure \ref{fig:fit_example} shows an example fit. 
In the top panel, the posterior best fit spectrum (orange) provides a visibly good fit to the observed PSB-region-stacked MaNGA spectrum (black), as is also seen by the small and normally distributed fitting residuals in the middle panel and the near unity reduced chi-squared.
The posterior spectrum contains the median physical model spectrum (blue) and the additive GP noise component (bottom panel). 
The latter does not exhibit an overall slope and has amplitude comparable to the observational uncertainty scaled by the posterior median uncorrelated noise amplitude (dark blue dashed line), which are signs of a well behaved model.
In the top panel, the physical spectrum is further separated to show dust-attenuated light contributions from the starburst (lime) and the older components (red). The split is placed at the time of minimum SFR between the old population and the starburst. This PSB has a burst mass fraction of $f_\mathrm{burst}=0.24^{+0.05}_{-0.03}$, and burst age of $t_\mathrm{burst}=0.61^{+0.12}_{-0.03}\;$Gyr, leading to a light contribution from the starburst that dominates marginally over the more massive older population in the red end of the optical spectrum, but more significantly at bluer wavelengths.

\begin{table*}
    \centering
    \caption{Posterior estimated properties of 50 PSBs from the spectral fitting of stacked MaNGA spaxels. 5 PSBs marked by dashes were poorly fit and are not considered in further analysis. Columns are (1) MaNGA Plate-IFU,
    (2) stellar mass within the stacked PSB spaxels,
    (3) ISM dust attenuation at 5500\AA\ ($V$ band),
    (4) the $\log_{10}$ SFR within the stacked PSB spaxels averaged over the last 100\,Myr,
    (5) time since the peak of the starburst,
    (6) fraction of mass formed during the starburst,
    (7) SFR halving timescale of the starburst,
    (8) stellar metallicity before the burst,
    (9) stellar metallicity during and after the burst,
    (10) change in metallicity,
    and (11) reduced chi-squared value of the maximum likelihood posterior sample spectrum. 
    The full table is available as supplementary online material.}
    \begin{tabular}{lllllllllll}
        \hline
        Plate-IFU & $\log_{10}$ & $A_V$ (mag) & $\log_{10}\mathrm{SFR_{100Myr}}$ & $t_\mathrm{burst}$ (Gyr) & \multirow{2}{*}{$f_\mathrm{burst}$ (6)} & $\tau_{1/2}$ (Myr) & $Z_{\rm{old}}/Z_\odot$ & $Z_{\rm{burst}}/Z_\odot$ & $Z_{\rm{diff}}/Z_\odot$ & $\chi^2_\nu$\\
        (1) & $M_{*,\mathrm{PSB}/M_\odot}$ (2) & (3) & ($\mathrm{M_\odot\; yr^{-1}}$) (4) & (5) &  & (7) & (8) & (9) & (10) & (11)\\
        \hline
        7961-1901  & $9.75^{+0.03}_{-0.03}$    & $0.85^{+0.03}_{-0.03}$    & $-0.65^{+0.07}_{-0.06}$   & $1.69^{+0.09}_{-0.07}$    & $0.81^{+0.11}_{-0.12}$    & $483^{+56}_{-35}$         & $1.59^{+1.07}_{-0.76}$    & $0.95^{+0.14}_{-0.15}$    & $-0.66^{+0.92}_{-1.15}$   & 0.950  \\
        7964-1902  & $9.19^{+0.05}_{-0.04}$    & $1.27^{+0.04}_{-0.04}$    & $-1.45^{+0.12}_{-0.14}$   & $1.86^{+0.29}_{-0.23}$    & $0.61^{+0.26}_{-0.19}$    & $499^{+112}_{-72}$        & $1.00^{+0.72}_{-0.40}$    & $1.27^{+0.26}_{-0.22}$    & $0.27^{+0.66}_{-0.92}$    & 0.965  \\
        7965-1902  & --                        & --                        & --                        & --                        & --                        & --                        & --                        & --                        & --                        & 0.878  \\
        8080-3702  & --                        & --                        & --                        & --                        & --                        & --                        & --                        & --                        & --                        & 0.928  \\
        8081-3702  & $8.76^{+0.06}_{-0.07}$    & $0.45^{+0.10}_{-0.08}$    & $-1.02^{+0.08}_{-0.08}$   & $0.93^{+0.14}_{-0.12}$    & $0.35^{+0.10}_{-0.08}$    & $535^{+167}_{-110}$       & $0.22^{+0.06}_{-0.09}$    & $3.19^{+0.22}_{-0.40}$    & $2.96^{+0.22}_{-0.35}$    & 0.971  \\
        \multicolumn{11}{c}{$\ldots$} \\
        \hline
    \end{tabular}
    \label{tab:results}
\end{table*}

\subsection{Most PSBs increase in stellar metallicity during starbursts} \label{sec:results1}
\begin{figure*}
    \centering
    \includegraphics[width=0.9\textwidth]{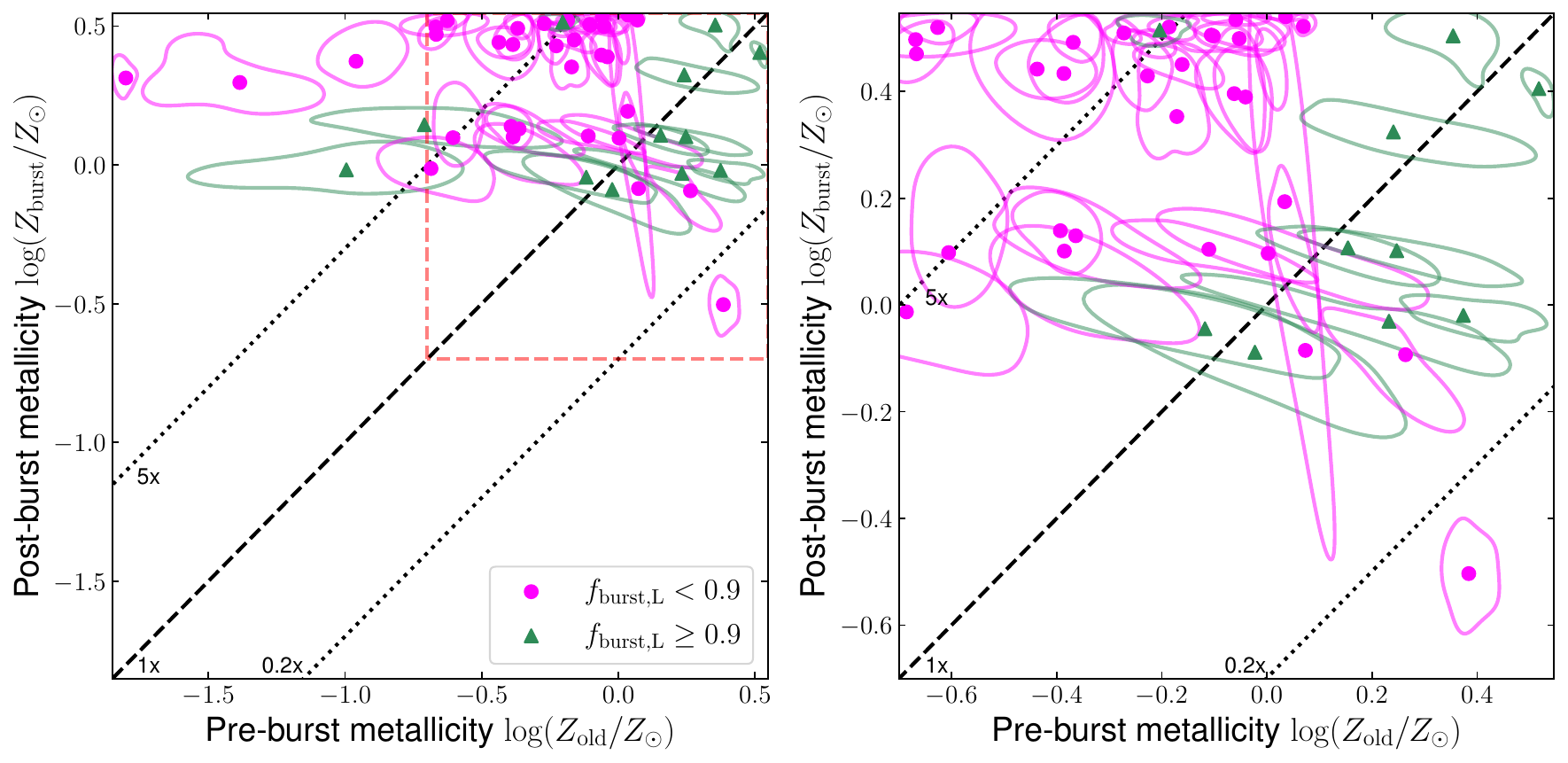}
    \caption{Pre-burst and post-burst median posterior stellar metallicities of PSB regions in MaNGA galaxies. The right panel is a zoomed-in view of the region bound by red dashed lines in the left panel. PSBs with a less dominating burst light fraction (posterior median $f_\mathrm{burst,L}<0.9$) are shown in magenta dots, otherwise in dark green triangles. The contours correspond to $1\sigma$ regions (enclosing the top 39.3\% of marginalised posterior probability), highlighting estimation uncertainties and degeneracies. The dashed black diagonal line marks constant stellar metallicity, while the dotted lines mark a $5\times$ and $0.2\times$ change in metallicity. Most PSB regions are found to increase in stellar metallicity during the starbursts.}
    \label{fig:zmet_old_vs_zmet_burst}
\end{figure*}

In Figure \ref{fig:zmet_old_vs_zmet_burst} we present the fitted posterior median metallicity levels before and after the starburst with $1\sigma$ contours to indicate posterior uncertainties.
Most PSBs (31/45, 69\%) lie above the diagonal constant metallicity line at $>1\sigma$ (20/45, 44\% at $>3\sigma$), indicating these PSBs experienced a significant increase in stellar metallicity during the starburst, many of which increased to $5\times$ the original metallicity (galaxies lying above the upper dotted line). 
A smaller fraction (4/45, 9\%) of PSBs are found to instead drop in metallicity at $>1\sigma$ (1/45, 2\% at $>3\sigma$), while the remaining portion (10/45, 22\%) have constant metallicity within the $1\sigma$ errors (24/45, 53\% within $3\sigma$).

Since estimating properties of the older stellar population is usually more challenging, the pre-burst stellar metallicity tends to have larger uncertainty. 
This uncertainty further increases where PSBs are found to have high burst light fractions ($f_\mathrm{burst,L}$) due to heavy outshining of the older population's light contribution.
We have therefore separated the sample at $f_\mathrm{burst,L}=0.9$ (calculated by integrating over the full fitted wavelength range) in Figure \ref{fig:zmet_old_vs_zmet_burst}.
The objects with high burst light fraction are seen to cluster around the line of constant metallicity (thick dashed line) but with large uncertainty, consistent with no change in metallicity being the a priori assumption given by the two-step metallicity model with independent and identical priors on metallicities.
Excluding these 13, the proportion of PSBs that experience a net positive change in stellar metallicity at $>1\sigma$ is 82\% (27/33).

\subsection{A recovered mass-metallicity relation, both before and after starburst}
\begin{figure*}
    \centering
    \includegraphics[width=0.95\textwidth]{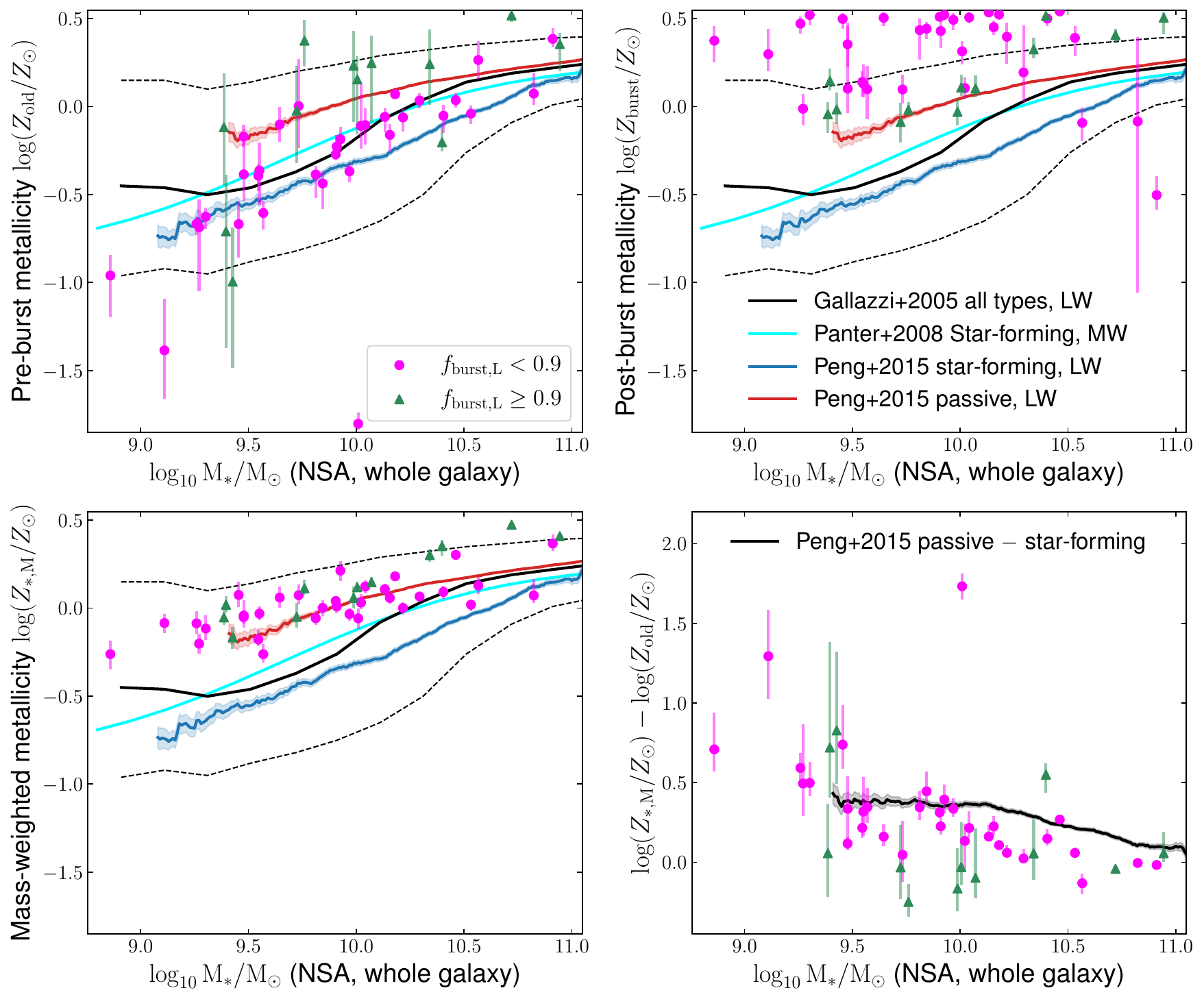}
    \caption{Stellar mass-metallicity relations of PSB regions both before (upper left), during the starburst (upper right) and overall mass-weighted (bottom left). PSBs with a less dominating burst light fraction (posterior median $f_\mathrm{burst,L}<0.9$) and a higher burst light fraction (posterior median $f_\mathrm{burst,L}\geq 0.9$) are marked with magenta dots and dark green triangles, respectively. All values are plotted against the estimated stellar mass of the whole galaxy from the NSA catalogue. Stellar mass-metallicity relations from the literature are also plotted for comparison, as indicated in the legends. The dashed black lines mark the 16th and 84th percentiles from \protect\cite{gallazzi2005}. The bottom right panel compares the difference between overall mass-weighted and pre-burst metallicity and the difference between passive and star-forming relations from the literature. The PSB pre-burst metallicity are found to agree with the \protect\cite{peng2015} star-forming relation, while the overall mass-weighted metallicity agrees with the \protect\cite{peng2015} passive relation, suggesting the PSB phase can explain the wide gap between the two literature relations.}
    \label{fig:zmet_vs_Mstar}
\end{figure*}

As reviewed in Section \ref{sec:intro}, the mass-metallicity (MZ) relation  has been used in the literature to infer the dominant quenching timescales that build up the red-sequence. 
It is interesting to compare the stellar mass and metallicity properties of our sample of PSBs to the MZ relation of local star-forming and quiescent galaxies, as the nature of PSBs being found soon after rapid quenching can provide insight into the impact of these quenching processes on chemical evolution.

The top left panel of Figure \ref{fig:zmet_vs_Mstar} shows our PSB sample's stellar mass-metallicity relation before the starburst occurred. 
Also shown are the MZ relations from three studies of local galaxies: star-forming SDSS galaxies, mass-weighted \citep{panter2008}; SDSS, all types, light-weighted \citep{gallazzi2005}; star-forming and passive SDSS galaxies, light-weighted \citep{peng2015}\footnote{We note that the \cite{peng2015} metallicity estimates are from \cite{gallazzi2005}, but split into star-forming and passive populations.}. 
Our PSBs broadly follow the known MZ relation where metallicity increases with mass, especially when we consider only the more reliable lower burst light fraction galaxies (magenta dots). 
This indicates that prior to the starburst, the PSB progenitors are consistent with being drawn from the underlying star-forming population, exhibiting no atypical chemical properties. 

The top right panel of Figure \ref{fig:zmet_vs_Mstar} shows our PSB sample's stellar metallicity during the starburst, which shows no observable correlations with total stellar mass, suggesting starbursts disrupt the MZ relation. 
In the bottom left panel, we show the overall mass-weighted stellar metallicity of the PSBs, which exhibit a remarkable agreement with the light-weighted mass-metallicity relation of passive galaxies from \cite{peng2015}. 
It is remarkable since local PSBs, being only recently quenched quiescent galaxies, might not be expected to be representative of the galaxy population at $z=0$. 
The difference between mass-weighted and light-weighted MZ relations should not affect our conclusions, since the difference is minor for quiescent galaxies \citep{trussler2020}.

In the bottom right panel of Figure \ref{fig:zmet_vs_Mstar} we compare the difference between overall mass-weighted metallicity and the pre-burst metallicity, with the difference between passive and star-forming galaxies from \cite{peng2015}. In both cases the differences decrease with increasing $M_*$. The matching trends point towards the PSB phase as a valid process that can create the large gap found between the star-forming and passive MZ relations reported in the literature. 
The implications are discussed in Section \ref{sec:discussion}.

\section{Discussion}\label{sec:discussion}
Our results show that most PSBs in our sample underwent an increase in stellar metallicity during the starburst phase, some very substantially. This indicates that the effect of stellar enrichment from the rapid recycling of gas during multiple rapid generations of star formation usually outweighs the combined effects of metal dilution from gas inflow and metal removal via outflow. In this section, we draw together studies from the literature to explain the metallicity changes we observe in PSB galaxies, and discuss the implications of our results for the role of PSB galaxies in galaxy evolution more generally. 

\subsection{On the implications of our results for the origin of PSBs}

Galaxy mergers have been a popular suggested trigger for local PSBs, supported by the large fraction of faint tidal features or companion galaxies \citep{zabludoff1996,chang2001}, high shape asymmetry \citep{pawlik2016,wilkinson2022}, neural network determined post-merger classification \citep{wilkinson2022} and unusual features in high spatial resolution images \citep{sazonova2021}. Recent mergers also exhibit a higher fractions of PSBs than non-merging galaxies \citep{ellison2022,li2023}. Observationally, a lowering of gas-phase metallicity is observed in both starbursts and pairs of interacting galaxies \citep{kewley2006,rupke2008,ellison2008}, apparently in contradiction to the results in this study. We explore this further below.

Simulations demonstrate how the disruption of the gravitational potential caused by a galaxy merger leads to strong torques that can drive a rapid gas inflow to the central regions, compress it and fuel a strong starburst \citep{barnes1991,barnes1996}. Forward modelling of these simulations have consistently shown this to be a reliable way to create galaxies with PSB spectral features and morphologies \citep{bekki2005,wild2009,snyder2011,davis2019,pawlik2019,zheng2020}. Comparatively fewer studies have focused on the chemical evolution of galaxies during gas-rich merger-induced starbursts. Since the outer regions of a star-forming galaxy are typically more metal-poor \citep[e.g.][]{matteucci1989,zaritsky1994}, the inflow of substantial gas driven by the disrupted gravitational potential would lead to a net decrease in central stellar metallicity, so long as the impact of stellar enrichment from the starburst is sufficiently weak. 

Initial hydrodynamic simulation studies found that gas funnelling events initially decrease the central gas-phase metallicity by diluting the existing relatively high metallicity gas, smoothing the negative radial metallicity gradients common to most star-forming galaxies \citep{perez2006,rupke2010,lahen2018} in agreement with the lowered gas-phase metallicity observed in local interacting and starburst galaxies. \citet{torrey2012} conducted ensembles of SPH simulations of major merging pairs of star-forming galaxies, finding that the change in stellar metallicity during the resulting starburst depends on the gas fractions of the progenitor galaxies: progenitors with low gas mass fractions tend to decrease stellar metallicity due to strong dilution from inflowing metal-poor gas during the merger, while progenitors with higher gas mass fractions tend to increase in stellar metallicity due to the stronger starburst and greater stellar enrichment. \citet{perez2011} found that gas-rich mergers drive a net increase in gas-phase metallicity of comparable magnitude to \citet{torrey2012}, though mainly caused by rapid increases in SFR due to fragmentation of the discs before merger-driven inflow occurs.

On the other hand, the more modern simulated major mergers from \citet{zheng2020} used in this work produce metallicity increases with only a weak trend with gas mass fractions. Orbits that produce stronger starbursts induce stronger metallicity enhancements, and minor mergers require higher gas fractions to achieve the same strength of starburst and therefore metallicity enhancement. Results are sensitive to the AGN feedback prescription used, as this impacts the strength of the starburst. Evidently there is scope for further simulation work on the chemical evolution of galaxies during gas-rich merger induced starbursts, to understand the impact of resolution, code type, AGN and chemical evolution modelling on the final properties of the descendants. The further development of semi-analytic models \citep[e.g.][]{molero2023},  for the specific case of low redshift elliptical formation, may also prove fruitful. 

Gas-rich galaxy mergers are not the only plausible cause of PSB galaxies in the local Universe. Ram-pressure stripping in dense galaxy clusters can compress the cold interstellar gas reservoir \citep{lee2017}, potentially leading to enhanced star formation in affected galaxies \citep{vulcani2020, roberts2022}, followed by rapid quenching \citep{poggianti2019,werle2022}. Although initially identified in clusters \citep{dressler1983}, it is important to note that PSBs are predominantly located in field environments \citep[e.g.][]{quintero2004,blake2004,goto2005, wild2009, pawlik2018}. The precise enhancement in the fraction of PSBs in dense clusters is still debated and may depend critically on redshift, stellar mass, and cluster and PSB selection methods \citep[e.g.][]{poggianti2009,vergani2010,vonderliden2010,socolovsky2018,paccagnella2019,wilkinson2021}.

Interestingly, lower stellar mass galaxies that are undergoing ram-pressure stripping are found to have elevated gas-phase metallicities compared to galaxies in both field and cluster environments of the same mass \citep{franchetto2020}, which might be a result of the increased stellar enrichment from ram-pressure compression without a significant dilution effect from metal-poor gas inflow. This would produce a rise in metallicity after a starburst, at least qualitatively similar to the metallicity increase seen in the majority of our PSB sample. 

To investigate further, we cross-matched our final sample of 45 well-fitted PSBs with the GEMA-VAC cluster catalogue \citep[][version DR17]{GEMA-VAC}, finding 11/45 to be members of rich clusters (member galaxies $>100$). This is around twice as high as a control sample (12.9\%) of galaxies from MaNGA DR17 matched in total stellar mass and $\mathrm{D_{4000}}$ stellar index at $1\mathrm{R_e}$ (MaNGA-Pipe3D, \citealt{manga-pipe3d1,manga-pipe3d2}; for $\mathrm{D_{4000}}$, see \citealt{bruzual1983}). However, we do not find any observable difference in the metallicity change and post-burst metallicity distributions of PSBs that are within rich clusters, compared to those that are not. Additionally, the PSBs and controls showed no significant difference in their distribution of local density as defined by the projected density to the fifth nearest neighbour (2-sample KS test $p=0.77$). Therefore, the importance of environmental processes are unclear from our present sample. 

While we find that the majority of PSBs in our sample have undergone significant increases in metallicity, a small number have experienced a metallicity drop. The most straightforward cause of a metallicity drop is strong inflow by metal-poor gas, with the inflow triggering a starburst that either produces not enough metals to counteract the effects of dilution, or the metals produced were preferentially expelled by outflows \citep{chisholm2018}. However, in Figure \ref{fig:zmet_vs_Mstar} the galaxies that experienced a metallicity drop are also the most massive. Since outflows are more effective at suppressing chemical enrichment at lower stellar mass \citep{tremonti2004,chisholm2018,maiolino2019}, the expelling of metals by outflows is likely not the main explanation for the metallicity decrease. We verified that there was no systematic correlation between the change in metallicity and size of the PSB regions, either in absolute size or relative to $R_e$, as might occur if the different patterns in metallicity evolution were caused by notably different merger types, or different processes entirely. The uncertainty in the simulations means that we cannot rule out mergers as a plausible trigger for either sets of PSBs.

\subsection{On the implications of our results for quenching pathways}\label{sec:discussion_quenching}

The evolution of stellar metallicity of galaxies has the potential to provide insight into the relative importance of different proposed quenching mechanisms. \citet{peng2015} measured the mass-metallicity (MZ) relation of star-forming and passive galaxies from SDSS, finding passive galaxies to be significantly more metal rich than star-forming galaxies with the same total stellar mass, with the gap widening for lower mass galaxies (see Fig.~\ref{fig:zmet_vs_Mstar}, lower right). The authors conclude the large MZ gap rejects quenching mechanisms that act on short timescales as a major contributor in quenching. They argue this is because rapid quenching would prevent significant increase in stellar metallicity as galaxies become passive, predicting little to no MZ gap, which is inconsistent with their observations. Instead, they favour slower mechanisms such as the strangulation of gas inflow, which allows for quenching galaxies to increase their metallicity from the star-forming MZ to the passive MZ through stellar enrichment, given enough time.  \citet{trussler2020} largely agrees with \citet{peng2015}, but additionally proposes ejection of gas through outflows to play a minor role in quenching. 

However, Figure \ref{fig:zmet_vs_Mstar} shows a good agreement between the metallicity evolution of our sample of PSBs and the star-forming-passive MZ gap. This indicates that the PSB phase, with relatively short starburst and rapid quenching that follows, is sufficient to provide the observed metallicity enhancement as galaxies move from the blue cloud onto the red sequence. Our results suggest long-term processes such as starvation are not the only viable pathways to explain the MZ gap as has previously been suggested. 

This result has global implications for galaxy evolution. Studies of the evolving stellar mass function of the red sequence have found it to grow rapidly during $z>1$ \citep[e.g.][]{ilbert2013,muzzin2013}, but the growth slowed or stalled by $z=1$ \citep{ilbert2013,rowlands2018a}. Therefore, a large fraction of the present-day red sequence likely quenched prior to $z=1$, Evidence from both observations \citep[e.g.][]{wild2009,wild2016,wild2020,whitaker2012,belli2019,taylor2023} and simulations \citep[e.g.][]{montero2019,zheng2022,walters2022} have suggested that PSBs and rapidly-quenched galaxies could contribute significantly to the growth of the red-sequence at $z>0.5$. Hence, a significant fraction of the local red sequence may have arrived there following a PSB phase. Making the leap that our local PSBs are likely undergoing similar chemical evolution to that experienced by PSBs at $z>1$, we thus conclude that short lived starbursts followed by rapid quenching might be a significant contributor to the observed MZ gap in local galaxies. 

Aside from slow-quenching through strangulation, other explanations for the gap in the local MZ relations between star-forming and passive galaxies have been proposed. 
Employing a ``leaky box" analytical model of chemical evolution, \cite{spitoni2017} modelled the two MZ relations and gas mass fractions of the star-forming galaxies to suggest that galaxies on the passive MZ relation might have assembled earlier, have shorter formation timescales and experienced weaker or less efficient outflows than star-forming galaxies of the same stellar mass. \cite{trussler2020} noted that a faster gas accretion timescale in passive galaxies could lead to a result similar to the shut-off of gas inflow in their strangulation hypothesis, supporting their slow-quenching picture. Our results do not directly provide evidence for or against the idea that the shorter assembly time of slow-quenching galaxies could lead to enhanced stellar metallicity once they enter quiescence, when compared to star-forming galaxies of the same stellar mass. However, our results do show that the above galaxy evolution pathway is not the only valid pathway resulting in passive galaxies that are more metal rich.

More recently, \cite{zibetti2022} drew connections to the local stellar mass surface density ($\mu_*$) to explain the observed gap in the MZ relations through investigating spatially resolved properties of galaxies in the CALIFA IFS survey. 
The study identified a single positive relation between stellar metallicity and $\mu_*$, and concluded that the higher global MZ relation in passive galaxies resulted simply from a lack of low mass surface density regions in these galaxies. These results are qualitatively similar to those found in an independent MaNGA sample from \cite{neumann2021}, and are in agreement with relationships found between $\mu_*$ and stellar metallicity \citep{barreraballesteros2017} and between $\mu_*$ and gas-phase metallicity \citep{gonzalezdelgado2014}.

The picture presented in \citet{zibetti2022} is consistent with our results and previous work on PSBs. The central starburst that precedes quenching in most PSBs likely enhances the local stellar mass surface density, leading to more compact structure \citep[e.g.][]{wellons2015}. Indeed, PSBs are generally found to be more compact than star-forming and quiescent galaxies of the same redshift and stellar mass \citep{yano2016,almaini2017,maltby2018,wu2018,chen2022,setton2022}. In addition, enhanced stellar surface mass density is expected to enhance local stellar metallicity through a combination of higher gravitational potential \citep{barone2018,barone2020} and a larger integral of the local SFH \citep{zibetti2022}, but the relationship's dependence on the shape of the SFH is unclear. Combining with our results showing that the PSB phase leads to a significant increase in stellar metallicity in most galaxies, a period of central starburst followed by rapid quenching can lead to the different distributions in stellar surface mass density and local stellar metallicity between younger and older regions found in \cite{zibetti2022}.

\subsection{On the implications of our results for the chemical evolution of starbursts}

Previous detailed theoretical work on the impact of bursty star formation on metallicity, and chemical abundance patterns more generally, has focused on local and relatively weak fluctuations in star formation rate, as might have occurred within regions of the Milky Way. On small scales, such as within the solar neighbourhood or within dwarf galaxies such as the Magellanic clouds, periods of increased efficiency of star formation (i.e. $\epsilon=\mathrm{SFR/M_{gas}}$) will lead to an increase in metallicity due to gas recycling and stellar enrichment \citep[e.g.][]{weinberg2017,johnson2019}. However, on global galaxy-wide scales, evidence for substantial enhancements in star formation efficiency in starbursts is still unclear, with inferred differences potentially driven by uncertainties in which CO-to-H$_2$ conversion factor to assume in different galactic environments (see \citealt{kennicutt_evans_2012} for a review, and \citealt{tacconi2018} for a summary of relevant results). 

We might expect the super-solar metallicity starbursts that we infer occurred in the recent past history of our PSB sample to be visible in analyses of the gas-phase mass-metallicity relation. The SFR dependence of the mass-metallicity relation for star-forming galaxies has been much debated in the past decade. \citet{barreraballesteros2017} use a sample of spatially resolved MaNGA galaxies to argue for no dependence of the MZ relation on star formation rate, and in particular there is no noticeable increase in metallicity at high sSFR in their data. However, our sample will represent $<1\%$ of the star forming population so will not be captured in large numbers in blind surveys. Previous studies have suggested extreme LIRGs or ULIRGs in the local Universe as progenitors to local PSBs, with LIRGs and ULIRGs having similarly low number densities \citep{hopkins2008,cales2011,french2015,pawlik2018}. The metallicity of such extreme starbursts is very difficult to estimate due to dust obscuration. A recent study by \cite{chartab2022} used mid IR strong line metallicity diagnostics to show that gas in local ULIRGs is not metal deficient as previously reported using standard optical line diagnostics. The difference arises due to dust obscuring the more metal enhanced star forming regions, and places ULIRGs firmly on the local MZ relation. Further work is clearly needed to verify whether super solar gas can be identified robustly in extreme starburst regions of local galaxies.

We searched for correlations between the stellar metallicity evolution and SFH in our PSB sample, which could further elucidate any relations between starburst properties and chemical evolution. However, the potential for sample selection effects to impact observed relations made it difficult to draw firm conclusions, and we therefore leave this to future work.

\subsection{Caveats and additional checks}\label{sec:discussion_caveats}
There are a number of caveats that are worth keeping in mind with regards to our study. 
The most important are the fact that we fit only the spatially resolved pre-selected PSB regions of the galaxies, and selection effects imposed by our selection of PSB spaxels.

We chose to fit only the PSB regions in the galaxies in order to simplify the SFH of the integrated spectrum, improving the accuracy of our results for these regions. 
By selection these regions are centrally located, and therefore represent the majority of light and mass in the galaxy, but some are more inclusive of the entire galaxy than others. 
Systematic correlations between the spatial extent of the PSB regions and a number of fitting galaxy properties are found in our sample. 
In particular, galaxies with larger PSB regions tend to have lower burst mass fractions but more rapid quenching, while those with smaller PSB regions have an even spread across the whole prior range. 
Further work is needed to explore the relation of these regions to the wider galaxy, and whether there are correlations between chemical evolution and the PSB spatial extent, particularly in the context of local stellar mass surface density as discussed in Section \ref{sec:discussion_quenching}.

To consider selection effects introduced by our PSB classification scheme (Section \ref{sec:data}), we calculate the theoretical selection fraction of galaxy models within our prior space as a function of parameters of interest. This is done by first randomly drawing $10^6$ mock galaxy models from the assumed SFH and metallicity priors, constructing mock spectra and measuring the spectral features $\mathrm{H\delta_A}$ and $\mathrm{W(H\alpha)}$. The fraction of mocks classified as PSB through our classification scheme is taken as the theoretical selection fraction. Although we found slight selection trends in a variety of physical properties (e.g. both older, weaker bursts and younger, slower decay bursts are less likely to be selected as PSBs), they were not consistent with causing any of the metallicity results presented here.

Minor degeneracies between SFH properties and post-burst metallicity are observed but do not qualitatively affect our results. We observed that for around half of the sample, post-burst metallicity is degenerate to a small extent with burst age ($t_\mathrm{burst}$) and burst mass fraction ($f_\mathrm{burst}$), where fitted models with an earlier, stronger starburst but lower post-burst metallicity, or models with a later, weaker starburst but higher post-burst metallicity have high posterior probabilities. However, the degeneracy is weak and are accounted for in the uncertainties of metallicity estimations shown in Figures \ref{fig:zmet_old_vs_zmet_burst} (contours) and \ref{fig:zmet_vs_Mstar} (error bars). Therefore, these degeneracies do not qualitative affect our conclusions. We do not see any degeneracies between any metallicity measurements with dust properties, velocity dispersion and redshift.

While alpha enhanced SSPs are available for older stellar populations \citep[e.g. ALF,][]{conroy2012,conroy2018}, and have been directly compared to \textsc{Bagpipes} \citep{carnall2022}, these models are not suitable for young or intermediate age ($\lesssim1\;$Gyr) stellar populations as found in our galaxies. 
The stellar population synthesis models we used do not include alpha-enhanced SSPs, and in this work we rely on the GP noise component to model out systematic uncertainties from spectral features related to alpha elements. 
To investigate this further we compared Mg and Fe absorption line indices measured from the stacked observed spectrum with those measured from our best-fit physical models (without GP noise), for all 45 well-fitted PSBs. We found that [$\alpha$/Fe] sensitive indices were substantially different in some cases, while [Fe/H] sensitive indices were relatively unchanged. This provides confidence that our metallicity estimates are not impacted by the lack of alpha enhancement in our models. 
In addition, we observe that PSBs with the strongest measured alpha enhancements were fitted with stronger GP noise corrections in the Mg absorption features, indicating that the GP noise component is effectively removing the model-data mismatch stemming from alpha element spectral features.
We conclude that the lack of alpha enhanced SSPs do not affect our measurements of stellar metallicity.

We note that our stellar metallicity estimates are limited by the metallicities of the stellar population synthesis models, which has an upper limit of $Z_*/Z_\odot\sim 3.52$. A number of our PSB's post-burst metallicity estimates are close to this upper limit (see Figure \ref{fig:zmet_old_vs_zmet_burst}). Although this is a potential issue for the accuracy of properties estimated for these PSBs, since the upper limit act to restrict the magnitude of metallicity increase during the starburst and subsequent rapid-quenching, our metallicity results cannot be caused by this limit.

As shown in Figure \ref{fig:masking}, we do not apply any cut on inclination during sample selection and both edge-on and face-on PSBs are included in our sample. 
To verify the effect of inclination on our results, we extract the 2D Sersic fit axis ratio (b/a) from the NSA catalogue \citep{blanton2011}, and found insignificant systematic correlations with our fitted galaxy properties in all cases ($p>0.05$, Spearman ranked correlation test).

\section{Summary and conclusions}\label{sec:conclusions}

Through selecting and stacking the PSB regions of 50 central PSB galaxies from the MaNGA integral field spectrograph survey, we fit the resulting high $\mathrm{SNR}>100$ stacked spectra with the Bayesian spectral energy density fitting code \textsc{Bagpipes}.
Taking inspiration from a suite of binary gas-rich merger simulations that created mock PSBs, we implemented a two-step metallicity evolution model where stars formed before and during the starburst are allowed independent metallicities. 
We reduced the computational time to fit the high SNR spectra by a factor of 100, by replacing the original GP kernel used in \textsc{Bagpipes} with a  stochastically-driven damped simple harmonic oscillator (SHOTerm), implemented through the \texttt{celerite2} code.
After careful verification of our fitting procedure through  ensembles of ``self-consistent” and simulation-based parameter recovery tests, we applied our model to the stacked spectra of MaNGA PSB regions to obtain 45 well-fitted results, where for the first time, the metallicity evolution of PSB galaxies with rapid SFH changes can be directly measured.
Our results lead to the following main conclusions:
\begin{enumerate}
    \item A majority (31/45, 69\%) of the PSB regions of galaxies formed significantly more metal-rich stars during the starburst than before (average increase $=0.8\;$dex with standard deviation $=0.4\;$dex), while a smaller number of PSB regions formed stars of equal or lower metallicity (Figure \ref{fig:zmet_old_vs_zmet_burst}). This suggests mechanisms that substantially raise stellar metallicity play important roles in the origin of PSBs: the effects of metal enrichment through stellar recycling outweigh those from dilution by gas inflow and metal removal by outflows.
    
    \item This  rise in metallicity during the starburst is consistent with simulations of gas rich mergers, agreeing with previous results that mergers are the leading cause of low redshift PSBs. However, we note that there is some disagreement on the impact of mergers on chemical enrichment in simulations, and more work needs to be done to corroborate the results from the \citet{zheng2020} simulations used here.
    
    \item A good agreement is found between the PSBs' pre-burst metallicity and star-forming mass-metallicity relations from the literature (Figure \ref{fig:zmet_vs_Mstar}, top left). This is consistent with PSBs being drawn from the underlying population of star-forming disk galaxies as expected.

    \item The PSBs' final mass-weighted mass-metallicity relation matches the local passive mass-metallicity relation. This suggests that the stellar metallicity evolution caused by rapid quenching following a starburst is entirely consistent with the observed gap in the stellar mass-metallicity relations between local star-forming and passive galaxies. Our results further validate the idea that rapid quenching following a starburst phase may be an important contributing pathway to the formation of the local quiescent galaxy population. 
    
\end{enumerate}

In this study we have focused on galaxies with central PSB features. 
Further work will be required to understand the importance of these features' spatial extent and how they compare to galaxies with other PSB spatial distributions \citep[e.g. ringed and irregular PSBs,][]{chen2019}.
The measurement of alpha enhancement in PSBs can allow for more precise timing of their starburst and quenching. 
Although difficult to obtain for recently quenched systems, alpha enhancement might be detectable in PSBs with older starbursts, for instance through the methods of \cite{conroy2018}. 
Lastly, further simulation work on the chemical evolution of galaxies during starbursts and rapid quenching is required, to understand the effects of AGN, shocks, stellar feedback, mergers/interactions and environments on chemical evolution.

\section*{Acknowledgements}
We thank the Referee for suggestions that improved the paper. We thank Dan Foreman-Mackey, Kartheik Iyer and Joel Leja for their assistance in using \texttt{celerite2}, \texttt{Dense Basis}, and \texttt{Prospector}, respectively. We thank Justus Neumann and Yingjie Peng for providing data. We also thank Justin Otter, Kate Rowlands and Omar Almaini and others in the UDS/PSB collaboration for useful discussions and insightful feedback. We thank Natalia Lah{\'e}n for feedback on the manuscript. We thank Laith Taj Aldeen for verification of fitting methods. VW and NFB acknowledge support from STFC consolidated grant ST/V000861/1. P.H.J. acknowledges support from the European Research Council via ERC Consolidator Grant KETJU (no. 818930). H-H.L thanks Alfie Russell and Sahyadri Krishna for assistance in language and phrasing.

Funding for the Sloan Digital Sky 
Survey IV has been provided by the 
Alfred P. Sloan Foundation, the U.S. 
Department of Energy Office of 
Science, and the Participating 
Institutions. 

SDSS-IV acknowledges support and 
resources from the Center for High 
Performance Computing  at the 
University of Utah. The SDSS 
website is www.sdss4.org.

SDSS-IV is managed by the 
Astrophysical Research Consortium 
for the Participating Institutions 
of the SDSS Collaboration including 
the Brazilian Participation Group, 
the Carnegie Institution for Science, 
Carnegie Mellon University, Center for 
Astrophysics | Harvard \& 
Smithsonian, the Chilean Participation 
Group, the French Participation Group, 
Instituto de Astrof\'isica de 
Canarias, The Johns Hopkins 
University, Kavli Institute for the 
Physics and Mathematics of the 
Universe (IPMU) / University of 
Tokyo, the Korean Participation Group, 
Lawrence Berkeley National Laboratory, 
Leibniz Institut f\"ur Astrophysik 
Potsdam (AIP),  Max-Planck-Institut 
f\"ur Astronomie (MPIA Heidelberg), 
Max-Planck-Institut f\"ur 
Astrophysik (MPA Garching), 
Max-Planck-Institut f\"ur 
Extraterrestrische Physik (MPE), 
National Astronomical Observatories of 
China, New Mexico State University, 
New York University, University of 
Notre Dame, Observat\'ario 
Nacional / MCTI, The Ohio State 
University, Pennsylvania State 
University, Shanghai 
Astronomical Observatory, United 
Kingdom Participation Group, 
Universidad Nacional Aut\'onoma 
de M\'exico, University of Arizona, 
University of Colorado Boulder, 
University of Oxford, University of 
Portsmouth, University of Utah, 
University of Virginia, University 
of Washington, University of 
Wisconsin, Vanderbilt University, 
and Yale University.

\textit{Software:} \textsc{Astropy} \citep{astropy}, \textsc{Bagpipes} \citep{bagpipes2018,bagpipes2019}, \textsc{Celerite2} \citep{celerite,celerite2}, \textsc{Dense Basis} \citep{iyer2019}, \textsc{Marvin} \citep{marvin}, \textsc{Matplotlib} \citep{matplotlib}, \textsc{MultiNest} \citep{multinest}, \textsc{Numpy} \citep{numpy}, \textsc{pipes\_vis} \citep{pipes_vis}, \textsc{Prospector} \citep{johnson2021}, \textsc{pyMultiNest} \citep{pymultinest}, \textsc{Scipy} \citep{scipy}, \textsc{Seaborn} \citep{seaborn}

For the purpose of open access, the author has applied a Creative Commons Attribution (CC BY) licence to any Author Accepted Manuscript version arising.

\section*{Data Availability}
All utilised MaNGA data are publicly available at the SDSS database \url{https://www.sdss4.org/dr17/} or through \texttt{Marvin} at \url{https://dr17.sdss.org/marvin/}. The stacked spectra and posterior samples of all 50 fitted galaxies are available at \url{https://doi.org/10.17630/ac0b406c-1c59-41e6-8b73-026790a0c1ca}. Python scripts to recreate the figures in Section \ref{sec:results} are available at \url{https://github.com/HinLeung622/chemical_evolution_of_PSBs_scripts}.



\bibliographystyle{mnras}
\bibliography{biblist} 




\appendix

\section{Calibration of the SHOTerm kernal}\label{apx:GP}
\begin{figure*}
    \centering
    \includegraphics[width=\textwidth]{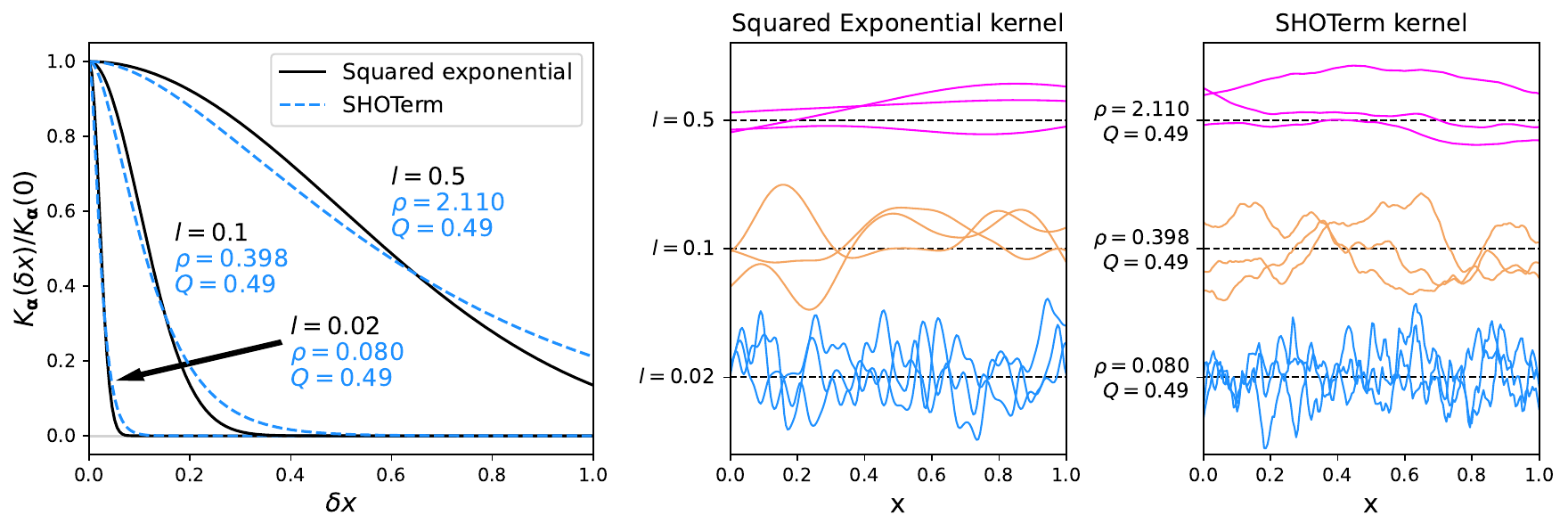}
    \caption{Comparing the GP kernels ``squared exponential'' and ``SHOTerm''. \textbf{Left}: Three auto-correlation curves from the squared exponential kernel of various lengthscales ($l$) (solid black) and their respective best fit auto-correlation curves from the SHOTerm kernel (dashed blue). The x-axis is the distance of separation in arbitrary units. The $l$ values of the squared exponential kernel are given in black next to each curve, while the $\rho$ and $Q$ values of the SHOTerm kernel are given in blue. SHOTerm favours a steeper initial decline in correlation, but has longer tails. \textbf{Centre}: 3 random realisations in arbitrary units drawing from each of the three squared exponential kernels in the left panel. \textbf{Right}: Similar random realisations drawn from each of the three best fit SHOTerm kernels. The realisations are qualitatively similar between the two kernels used, but the SHOTerm kernels display increased fluctuations at small lengthscales compared to the squared exponential kernels. This is due to the steeper initial decline in correlation.}
    \label{fig:GP_demo}
\end{figure*}

Here we detail how the SHOTerm kernal was calibrated to provide a qualitatively similar result to the squared exponential kernal successfully employed in \citet{bagpipes2019}. Firstly, squared exponential auto-correlation curves were obtained by selecting values of $l$ within its prior range \citep[as specified in][]{bagpipes2019}. Then, $\rho$, $\sigma$ and $Q$ values of the SHOTerm kernel that produce the best agreement between auto-correlation curves of the two kernels were computed using least squares. We found fixing $Q=0.49$, along with setting the period prior as $0.04<\rho<1.0$ and the standard deviation prior as $10^{-1}<\sigma<1$, both with shapes flat in $\log_{10}$ space (as listed in Table \ref{tab:priors}) to best approximate the original kernel's behaviours. 
In Figure \ref{fig:GP_demo}, we compare three examples of the squared exponential kernel of various $l$ and their corresponding best fit SHOTerm kernels. The figure's left panel compares the kernels' auto-correlation function, showing that SHOTerm favours a steeper initial decline in correlation, but has longer tails. As a result, when comparing random realisations drawn from the kernels in the central and right panels, the SHOTerm kernels display an increased level of fluctuations at small lengthscales. However, at large lengthscales the realisations are qualitatively similar.

\section{SPH Simulation parameter recovery without GP noise}\label{apx:noGP_test}
\begin{figure*}
    \centering
    \includegraphics[width=0.93\textwidth]{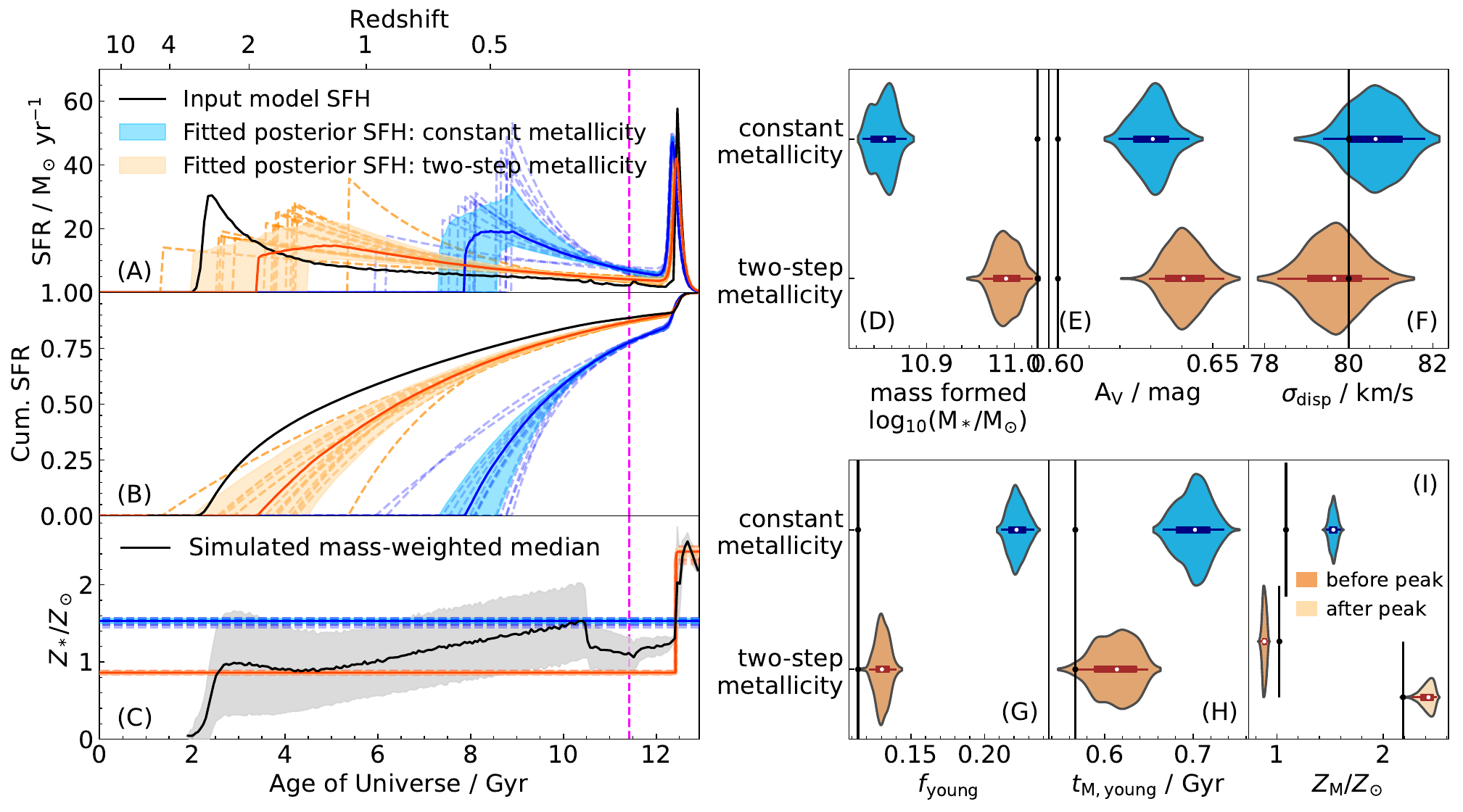}
    \caption{Simulation-based parameter recovery test without GP noise component, comparing the results of fitting the constant (blue) and two-step (orange) metallicity models to star particles generated from an SPH merger simulation \citep[from][see Figure \ref{fig:yirui_sim}]{zheng2020}. All symbols follow the same meaning as in Figure \ref{fig:particle_test}. The two-step metallicity model continues to out-perform the constant model even without the GP noise component.}
    \label{fig:particle_test_noGP}
\end{figure*}

\begin{figure*}
    \centering
    \includegraphics[width=\textwidth]{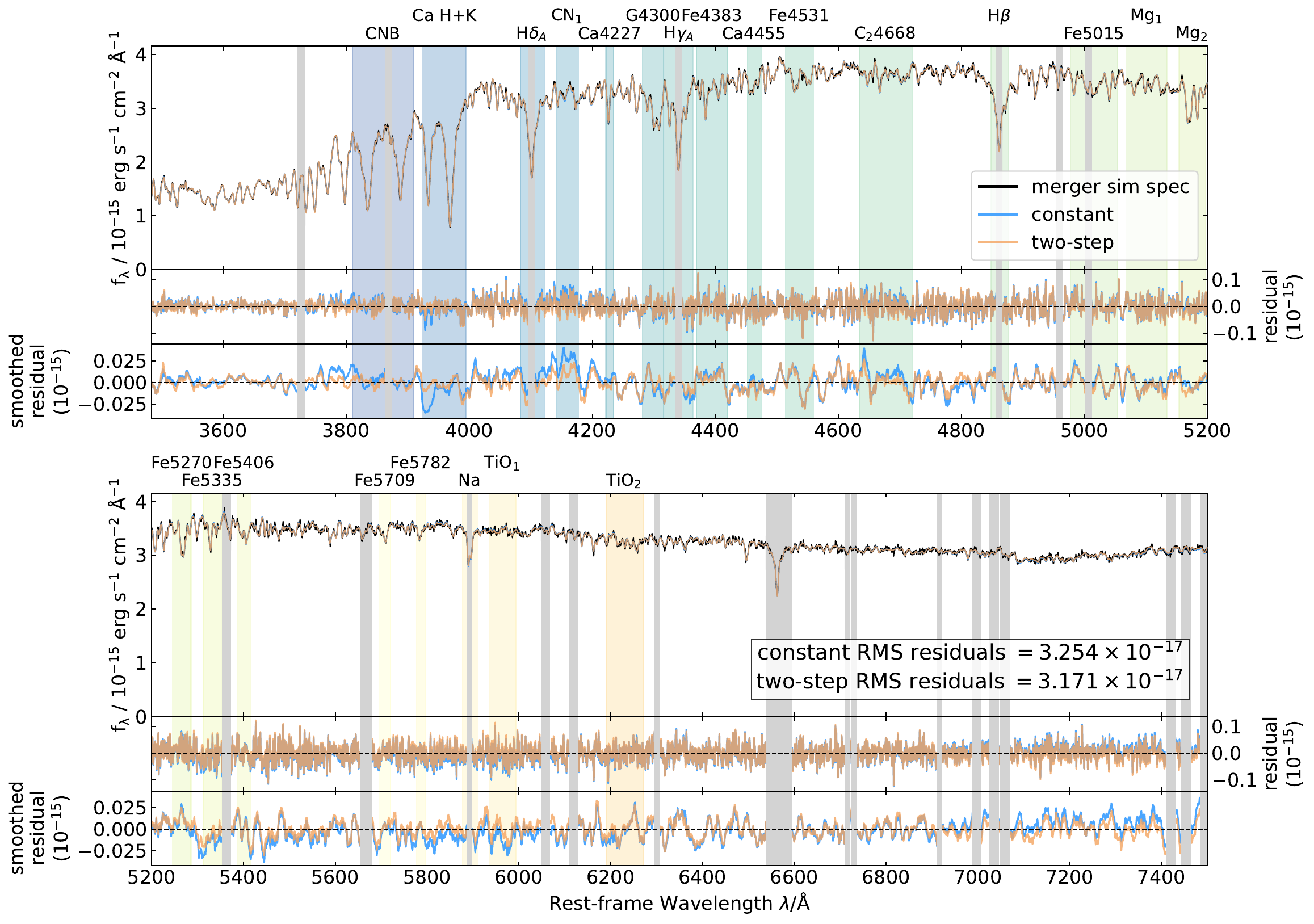}
    \caption{Comparing the spectral fitting performance of the constant (blue) and the two-step (orange) metallicity models applied to a mock PSB spectrum generated from an SPH merger simulation (Figure \ref{fig:particle_test_noGP}), without the GP noise component. All symbols follow the same meaning as in Figure \ref{fig:particle_test_spec}, except the GP noise panels are removed. The constant metallicity model remains a poorer fit at most metal indices. The residual at calcium H+K lines worsened.}
    \label{fig:particle_test_spec_noGP}
\end{figure*}

As seen in Figure \ref{fig:particle_test_spec}, the constant metallicity model's GP noise component led to a best fit model with a prominent slope, allowing the physical spectrum to be significantly bluer than the input mock spectrum. Therefore, to investigate the impact of the GP noise component on parameter recovery performance and minimizing residuals, we repeat the pair of tests with the two metallicity models fitting the same mock spectra constructed from the binary merger simulation, but without the GP noise component. Figures \ref{fig:particle_test_noGP} and \ref{fig:particle_test_spec_noGP} shows the no-GP counterparts to Figures \ref{fig:particle_test} and \ref{fig:particle_test_spec}.

Without the GP noise component, comparing Figures \ref{fig:particle_test_noGP} and \ref{fig:particle_test}, SFH and parameter recovery for both metallicity models declined, accompanied with increased uncertainties. However, the two-step metallicity model remains the more accurate model. Comparing Figures \ref{fig:particle_test_spec_noGP} and \ref{fig:particle_test_spec}, the two-step model remains the better fit at the metal-indicating calcium H+K lines, and the iron and magnesium indices. Additionally, the residual of the constant model worsened, most notably around the H+K lines. This leads to the larger difference in root mean square residuals between the two models. In conclusion, the two-step metallicity model's superior parameter recovery performance is not due to the GP noise component of the model.

\section{Fitted star-formation histories and metallicity levels}\label{apx:SFH}
\begin{figure*}
    \centering
    \includegraphics[width=0.9\textwidth]{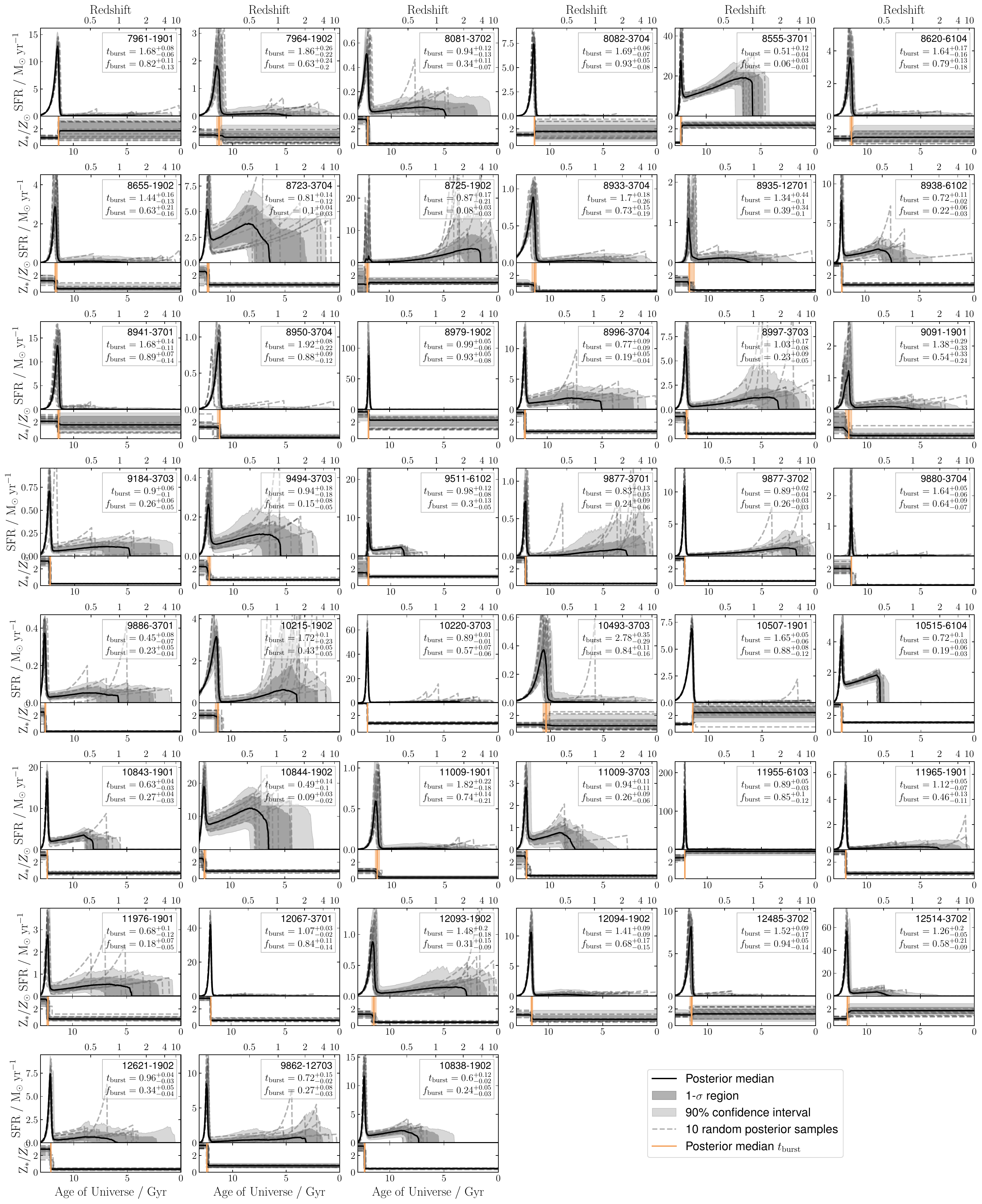}
    \caption{Fitted SFHs (top panel of each subplot) and metallicity evolution (bottom panel of each subplot) of 45/50 successfully fitted PSBs. The solid black lines, grey and light grey shaded regions mark the posterior median, the $1\sigma$ region and 90\% confidence interval of the fitted SFHs or metallicity histories. Ten random samples drawn from the posterior distributions are shown using grey dashed lines. The vertical orange lines and orange shaded regions mark the posterior median and $1\sigma$ region of the time of peak of starburst ($t_\mathrm{burst}$).}
    \label{fig:SFH_all}
\end{figure*}

Figure \ref{fig:SFH_all} shows the derived SFH's and metallicity evolution for the 45/50 successfully fitted galaxy in our sample. The spectral fitting methods used are described in Section~\ref{sec:fitting}. SFH parameters such as the time since burst, burst mass fraction, etc are listed in Table \ref{tab:results}. 

\bsp	
\label{lastpage}
\end{document}